\documentclass[reprint, amsmath,amssymb, aps,superscriptaddress,prx,floatfix,longbibliography]{revtex4-2} 
\usepackage{graphicx}
\usepackage{siunitx}
\usepackage[T1]{fontenc}
\usepackage[utf8]{inputenc}
\usepackage{xcolor}
\definecolor{codegreen}{rgb}{0,0.6,0}
\definecolor{codegray}{rgb}{0.5,0.5,0.5}
\definecolor{codepurple}{rgb}{0.58,0,0.82}
\definecolor{backcolour}{rgb}{0.95,0.95,0.92}
\definecolor{urlblue}{HTML}{007bff}

\usepackage{bm}
\usepackage{bbm}
\usepackage{orcidlink}
\usepackage{microtype}
\usepackage{enumitem}
\usepackage{braket}
\usepackage{mathtools}
\usepackage{soul}
\usepackage{ulem}

\usepackage{times} 

\usepackage[capitalise]{cleveref}
\input{macros.sty}

\hypersetup{
    colorlinks = true,
    linkcolor  = urlblue,
    citecolor  = urlblue,
    urlcolor   = urlblue,
}

\usepackage[mathlines]{lineno}

\begin{document}


\title{
Fractonic Fractional Quantum Hall Effect
}

\author{Justin Schirmann\orcidlink{0009-0007-7030-0155}
}
\thanks{These authors contributed equally}
\affiliation{\small Univ. Grenoble Alpes, CNRS, Grenoble INP, Institut Néel, 38000 Grenoble, France}

\author{Peru d'Ornellas\orcidlink{0000-0002-2349-0044}}
\thanks{These authors contributed equally}
\affiliation{\small Univ. Grenoble Alpes, CNRS, Grenoble INP, Institut Néel, 38000 Grenoble, France}

\author{Charles Stahl\orcidlink{0000-0002-9809-5575}}
\affiliation{\small Department of Physics, Stanford University, Stanford, CA 94305, USA}

\author{Adolfo G. Grushin\orcidlink{0000-0001-7678-7100}}
\affiliation{\small Univ. Grenoble Alpes, CNRS, Grenoble INP, Institut Néel, 38000 Grenoble, France}
\affiliation{Donostia International Physics Center (DIPC),
Paseo Manuel de Lardiz\'{a}bal 4, 20018, Donostia-San Sebasti\'{a}n, Spain}
\affiliation{IKERBASQUE, Basque Foundation for Science, Maria Diaz de Haro 3, 48013 Bilbao, Spain}%

\date{\today}

\begin{abstract}

In non-interacting systems, disorder can drive a trivial phase into a topological one. 
However little is known how to construct a fractional quantum Hall ground-state, a paradigmatic topologically ordered state, that exists both in crystalline and disordered lattices \textit{and} is qualitatively different to known topological phases.
Here, we propose a general method for building such a phase. 
This is done by coupling quantum wires placed aperiodically in real-space, where the spatial positioning allows us to tune the inter-wire couplings.
We call the emergent phase the Fractonic Fractional Quantum Hall Effect as it displays a rich interplay of fractional quantum Hall physics with fractonic constraints, formed by coupling differently-fractionalised wires into a globally gapped phase.
The ground state has an exponential degeneracy in system size, a signature of the emergence of fractons. 
It displays a rich phenomenology of excitations, which can either behave like anyons confined to move in one dimension (lineons), multiples of which can then hop between two wires (s-lineons) or be free to travel across the system (C-anyons), depending on the multiplicity.
Both the ground state degeneracy and mutual statistics are directly determined by the real-space positions of the wires, which can be disordered. 
Our method provides an analytically solvable pathway to non-crystalline fractional quantum Hall effects and fractonic theories in two-dimensions, examples of which were lacking.

\end{abstract}

\maketitle
\tableofcontents

\section{Introduction}

Topologically ordered phases encode information non-locally through long-range entanglement, a feature that can be exploited to engineer quantum computers robust to decoherence by braiding the underlying anyon excitations~\cite{Kitaev2003,Nayak2008}. Hence, finding and classifying topologically ordered phases is a central goal of condensed matter~\cite{wen2004quantum}.

The search for topological order is often guided by progress in crystalline systems, and builds on our understanding of non-interacting topological phases. As for topological band insulators, topological order
was shown to survive until the strength of disorder closes the mobility gap. For example, the plateaus of the fractional quantum Hall effect~\cite{Tsui1982}, a paradigmatic example of topological order~\cite{Laughlin1983}, shrink and disappear for sufficiently strong potential disorder~\cite{Sheng2003} and electric conductance can exhibit mesoscopic fluctuations~\cite{Kane1994,Kane1995,haldane_stability_1995,Kao1999,Yutushui2024}.
This would suggest that looking for topological order in a non-crystalline system is not a fruitful strategy.

However, there are several examples where strong disorder is not detrimental, and can even be beneficial, for topological order.
For instance, the toric code ground state~\cite{Kitaev2003}, another topologically ordered state, can be defined in any planar graph~\cite{Kitaev2003,Simon_TopologicalQuantum}. It is also possible to define fractional quantum Hall states with trial wave-functions on non-crystalline lattices, for example in fractals~\cite{Manna2020,Manna2022,Manna2023,Xikun2022,Jha_2023,Jaworowski2023} and quasicrystals \cite{Duncan2020}. 
 
More strikingly, disorder or non-crystallinity can help to tune into and control the properties of a topologically ordered state. For example, moir\'{e} heterostructures,  lacking strict lattice periodicity,  allow topological bands to flatten by rotating two monolayer materials with respect to one another. By exploiting this freedom, recent experiments have reported the anomalous fractional and integer quantum Hall effect in twisted MoTe$_2$~\cite{Cai2023,Zeng2023,Park2023,Xu2023,Kang2024,xu2025signatures} 
and rhombohedral multilayer graphene~\cite{Lu2024,Xie2024,Waters2024,Lu2025}, realizing the long-sought prediction of a fractional Hall effect without magnetic fields~\cite{Neupert2011,Tang2011,Sun2011,Sheng2011,Regnault2011}. 
Additionally, structural disorder or amorphisation can open a gap in Kitaev's honeycomb quantum spin-liquid phase leading to a topological, non-abelian chiral spin-liquid phase~\cite{Grushin2023,Cassella2023}. This success motivates the question we address in this work. 

While non-crystalline systems can realise topological order, can we go further and exploit non-crystallinity to define topologically ordered states with different features compared to their crystalline counterparts? Recent advances in non-crystalline, single particle topological states, which have been defined in several non-crystalline settings, including quasicrystals~\cite{Kraus:2012iqa,Tran:2015cj,Bandres:2016gx,Fuchs:2016hp,Huang2018,Fuchs:2018dd,Loring2019,Varjas2019,Chen2019,He2019,Duncan2020,Zilberberg2021,Hua2021,Jeon2022,Schirmann2024}, amorphous systems~\cite{agarwala2017,Mansha2017,Xiao2017,mitchell_amorphous_2018,Bourne:2018jr,Poyhonen2018,minarelli_engineering_2019,chern_topological_2019,mano_application_2019,corbae_evidence_2020,Costa:2019kc,marsal_topological_2020,Sahlberg2020,ivaki_criticality_2020,agarwala_higher-order_2020,Grushin2020,wang_structural-disorder-induced_2021,Corbae2021,focassio_structural_2021,Mitchell2021,spring_amorphous_2021,wang_structural_2022,marsal_obstructed_2022,Peng2022,uria-alvarez_deep_2022,spillage_2022,Corbae_2023_review,Cassella2023,Grushin2023,Manna2024,Ciocys2023, uria2024amorphization} and fractals~\cite{Brzezi2018,Pai2019,Iliasov2020,Fremling2020,Yang_fractal_2020,manna_noncrystalline_2022,Manna2022b,LI20222040,ZHENG20222069,Biesenthal2022,Ivaki2022,Manna2002_HOTI_fractal,Boquan2023,Salib2024,Lai2024,Canyellas2024,Li_fractal_2024,Lage_2025}, suggest that this can indeed be the case.

The goal of this paper is to propose a solvable model for a non-crystalline fractional quantum Hall phase with different properties compared to its crystalline counterpart. %
This is done using a system of coupled one-dimensional wires, a successful strategy for constructing solvable fractionalised topologically ordered phases~\cite{Emery2000,Vishwanath2001,Mukhopadhyay2001,Mukhopadhyay2001b,kane_fractional_2002,teo_luttinger_2014,tam_nondiagonal_2021,tam_coupled_2020,tam_toric_2022}.%

The coupled-wire construction has been used to construct a rich variety of anisotropic, yet still crystalline, many-body states. These include fractionalised topological phases~\cite{Clarke2013,Neupert2014,Klinovaja2014,oss14,sagi_non-abelian_2014,santos_fractional_2015,Cano2015,Klinovaja2015,Sagi2015,sagi_imprint_2015,Meng2015,meng_theory_2016,Mross2016,Iadecola2016,Fuji2016,Sagi2017,Kane2017,Park2018,Kane2018,fuji_quantum_2019,Laubscher2019,Bardyn2019,Imamura2019,Han2019,meng_coupled-wire_2020,crepel_microscopic_2020,Laubscher2020,Li_wire_2020,tam_nondiagonal_2021,Laubscher2021,Zhang2022,imamura_coupled_2019,Laubscher2023,Pinchenkova2025}, spin liquids~\cite{meng_coupled-wire_2015,Gorohovsky2015,Patel2016,Huang2016,Lecheminant2017,Pereira2018,Ferraz2019,Slagle2022,Mondal2023,gao2025triangularlatticemodelskalmeyerlaughlin} and fractonic states~\cite{Gabor2017,Leviatan2020,sullivan_fractonic_2021,May-Mann2022,fuji_bridging_2023,You2025}. 
More recently, the wire construction has also allowed us to analytically describe strongly correlated topological states in twisted moir\'{e} heterostructures, as a network of crossing coupled wires~\cite{Wu2019,Chou2019,Chen2020,Chou2021,Lee2021,Hsu2023,Biao2024,Shavit2024,Hu2024,wu_possible_2024}. 

Despite its success, the wire construction has been exclusively used to describe crystalline phases. A possible reason is that the couplings necessary for a fractionalised phase must be precisely chosen to satisfy charge and momentum conservation, while still retaining exact solubility. This places strong constraints on the arrangement of the wires and generally forbids disordered configurations from admitting an exact solution under bosonisation. One might then worry that any model with non-uniform spacing between the wires becomes unsolvable.

\begin{figure}[t]
    \centering
    \includegraphics{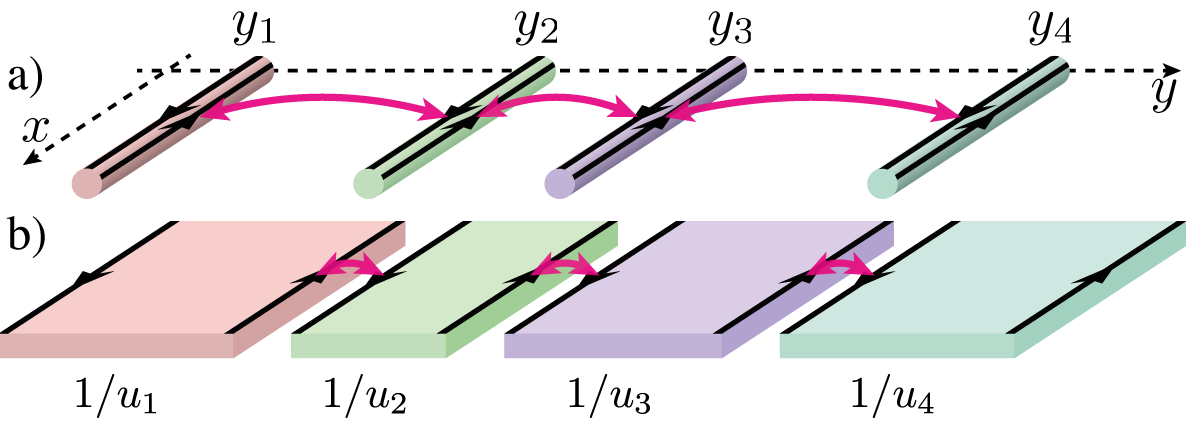}
    \caption{
    The mapping from (a) non-uniformly coupled wires to (b) coupled Hall droplets. Wires are placed at a set of (non-uniformly spaced) positions $\{y_j\}$ and coupled together to form a non-uniform fractional Hall phase. Couplings depend on the inter-wire separation, and so are themselves non-uniform. The left- and right-movers on each differently fractionalised wire can be mapped onto the left- and right-moving edge modes in each fractional Hall droplet.
    The couplings necessary to gap out the non-uniform arrangement of wires in (a) are identical to the couplings necessary to gap out the fractionalised gapless edge modes of adjacent fractional quantum Hall droplets in (b).}
    \label{fig:intro_droplets_wires}
\end{figure}

Here we show that this is not the case. By considering coupled wires whose real space configuration is disordered and non-uniform, 
shown in \cref{fig:intro_droplets_wires}a, we show that the problem remains solvable via bosonisation. We formalise the constraints on the disordered couplings under which the bosonic fields commute among themselves---thus retaining exact solvability.By tuning the spacing between wires we show that it is possible to induce hybridisation between \textit{differently fractionalised} quasiparticles between each pair of adjacent wires, effectively mixing different fractionalisaions to form a global phase with qualitatively new properties. This procedure is possible because one can gap out fractional quantum Hall edge states with different fractionalisations whenever the ratio of fractionalisations is a square number~\cite{haldane_stability_1995,santos_parafermionic_2017,may-mann_families_2019}, depicted in \cref{fig:intro_droplets_wires}b. 

We show that the non-crystalline fractional quantum Hall states constructed in this way display features distinct from known crystalline counterparts. Specifically, we find that our system has many features of a gapped fractonic phase. This is remarkable because it is a 2D construction \cite{Nandkishore_2019,sullivan_fractonic_2021,Pretko_2020} and currently no mechanism is known for producing fractons in 2D \cite{aasen_topological_2020}. We show that the system has a ground state degeneracy that scales exponentially with system size. Abelian anyonic quasiparticles live at the interface between each pair of wires and carry statistics that depend on the real-space arrangement of their two adjacent wires. Single quasiparticles are partially immobile, where no physical operator can transport a single fracton between wires---and so behave as lineons. We show that despite being immobile alone, bundles of these elementary lineons have less limited mobility, where the multiplicity of the composite defines its transport. We describe two such composite quasiparticles. Lineons may be combined to form an excitation that is able to travel up to the next wire and back (which we call a spread-lineon or s-lineon for short). S-lineons may then be combined to form a composite anyon with complete freedom to travel throughout the system, which we refer to as a C-anyon. The mutual statistics of these excitations depend on the real-space arrangement of the wires, and differ from that of the constituent fractional Hall regions.

The outline of this paper is as follows: In \cref{sec:the_model} we lay the foundations for our model, defining the wire construction and its reformulation in terms of bosonised operators. In \cref{sec:fractionalisation} we show how the degrees of freedom in a single wire may be re-expressed in a basis of fractionalised bosonic operators. Next, we show how couplings may be introduced between adjacent wires that gap out these fractionalised excitations, thus allowing the system of wires to host a single correlated phase of matter. Importantly, we see that it is possible to couple two adjacent wires even when each hosts differently fractionalised excitations, and discuss the conditions under which this coupling preserves exact solvability and is gapped. In \cref{sec:general_construction} we show how this constitutes a general construction of composite and disordered fractional phases, created by gluing many regions with different fractionalisation together. We study the properties of the quasiparticles (both fractonic and non-fractonic) that emerge, determining their charge, reduced mobility and braiding statistics. Finally, in \cref{sec:fractonic} we provide a proof of two fractonic properties. We show that the reduced fractonic mobility is completely robust---that is, no well-posed operator can be written down that transports quasibarticles in a way that breaks the fracton constraint. Finally, we prove that the ground state degeneracy of the system scales exponentially with the system size, leaving the stability of the degeneracy to future work.

\section{The Model}\label{sec:the_model}

\begin{figure}[t]
    \centering
    \includegraphics{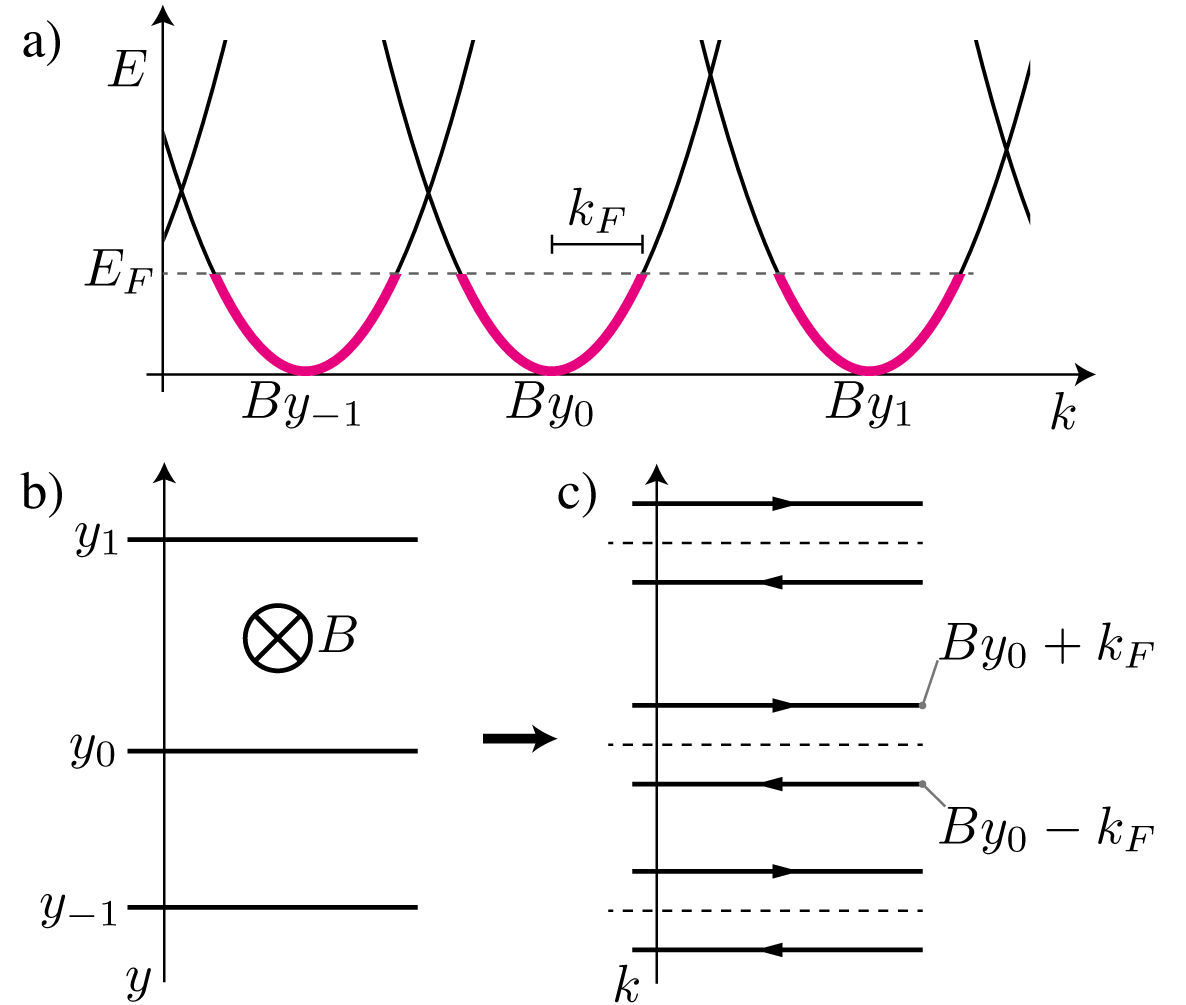}
    \caption{ 
    (a) Dispersion for three coupled wires placed at positions $y_j$ in a field $B$. Filled states are highlighted in magenta and the Fermi momentum is labelled with $k_F$. Note that wires here are not regularly spaced, so the centres of the parabolas are also not evenly spaced.
    (b) Real space diagram showing the positions of each wire, and the orientation of the applied field. 
    (c) The low-energy spectrum of each wire is effectively split into a set of left movers and right movers, shifted from the central $k$ of each wire by $k_F$. Dashed lines indicate the values of $By_j$.
    }
    \label{fig:wires_Fermi_level}
\end{figure}

We start by introducing the model, which will follow a familiar procedure to Refs.~\cite{kane_fractional_2002,teo_luttinger_2014,tam_nondiagonal_2021}. We construct an array of coupled fermionic wires, aligned in the $x$-direction, with the $j^{\textup{th}}$ wire placed at position $y_j$. Penetrating the plane is a magnetic field $B$, which we represent in the Landau gauge, $\textbf A = -B y \hat {\textbf x}$. On each wire sits a single species of fermion, represented with the creation operators $\psi_j(x)$ that satisfy
\begin{align}
    \{\psi_j^{\dagger}(x),\psi_k^{\pdag}(x')\} = 
    2\pi \delta_{jk}\delta(x-x').
\end{align}
The Hamiltonian for each uncoupled wire is given by
\begin{equation}
    \label{eqn:ferminic_single_wire}
    H_j = \int \frac{\mathrm{d} x}{2\pi} 
    \psi_j^{\dagger}(x)
    \frac{\left(-i\partial_x - By_j\right)^2}{2m}
    \psi_j^{\pdag}(x),
\end{equation}
which leads to a parabolic dispersion for the fermions on each wire, with the parabola shifted according to the wire's position, 
$E_j(k) = \left (k - B y_j\right )^2/2m$.
We impose a global Fermi level and corresponding Fermi momentum, $E_F = k_F/2m$, such that in each wire all momentum states that satisfy $|k_j - By_j| \leq k_F$ are filled, as shown in \cref{fig:wires_Fermi_level}a.

For a given Fermi level, the low-energy, long-wavelength properties of the system are defined only by the electrons close to $E_F$. Thus, we may construct an effective description by linearising the dispersion relation around these points. On each wire, the $\psi_j(x)$ fermion can be decomposed into a pair of $\psi^R_{j}(x)$ and $\psi^L_j(x)$ fermions, corresponding to right-movers and left-movers, by separating positive and negative momentum states. We thus derive the \textit{linearised Hamiltonian} for a given wire as
\begin{align}
    H_j^{\textup{Lin}} &= H_j^{R} +H_j^{L} , \label{eqn:linearised_ham}\\
    H_j^{R} &= \int \frac{\mathrm{d} x}{2\pi} v
    {\psi^R_{j}}^\dag(x) (- i\partial_x ) {\psi^R_{j}}^{\pdag}(x),\label{eqn:linearised_R} \\
    H_j^{L} &= \int \frac{\mathrm{d} x}{2\pi} v
    {\psi^L_{j}}^\dag(x) ( i\partial_x ) {\psi^L_{j}}^{\pdag}(x),\label{eqn:linearised_L}
\end{align}
with $v = \frac{k_F} m $. Each wire is effectively split into a pair of wires, each with unidirectional transport at a pair of momenta shifted by $\pm k_F$, as shown in \cref{fig:wires_Fermi_level}b. Note that \cref{eqn:linearised_L,eqn:linearised_R} should include a shift by the momentum at the Fermi level, $k_j^{p} = y_jB + (-1)^p k_F$, where $p\in \{R,L\}$. This shift is removed by redefining $\psi^a_j(x) \rightarrow e^{ik_j^{p} x}\psi^a_j(x)$.

\subsection{Allowed Couplings} \label{sec:allowed_couplings}

To realise an effective 2D phase, we must now introduce couplings between adjacent wires. A general coupling is expressed in terms of a set of parameters $s_j^{R}$ and $s_j^{L}$, that denote the number of right and left-movers created respectively on the $j^{\textup{th}}$ wire. Such a coupling may be written as
\begin{align}
\label{eqn:coupling_operator}
    \mathcal O_{\{s_j\}}(x) \propto 
    \prod_{j}
    {\psi^R_j}^{(s_j^{R})}(x) 
    {\psi^L_j}^{(s_j^{L})}(x) 
    + h.c.,
\end{align}
where we write ${\psi^{|s_j^p|}_j}^\dag$ when $s_j^{p} > 0$. Since fermionic operators square to zero, these terms must be written using a point-splitting prescription~\cite{von_delft_bosonization_1998}. The \textit{coupling Hamiltonian} is expressed as a sum over all such coupling terms.

\begin{figure}[t]
    \centering
    \includegraphics{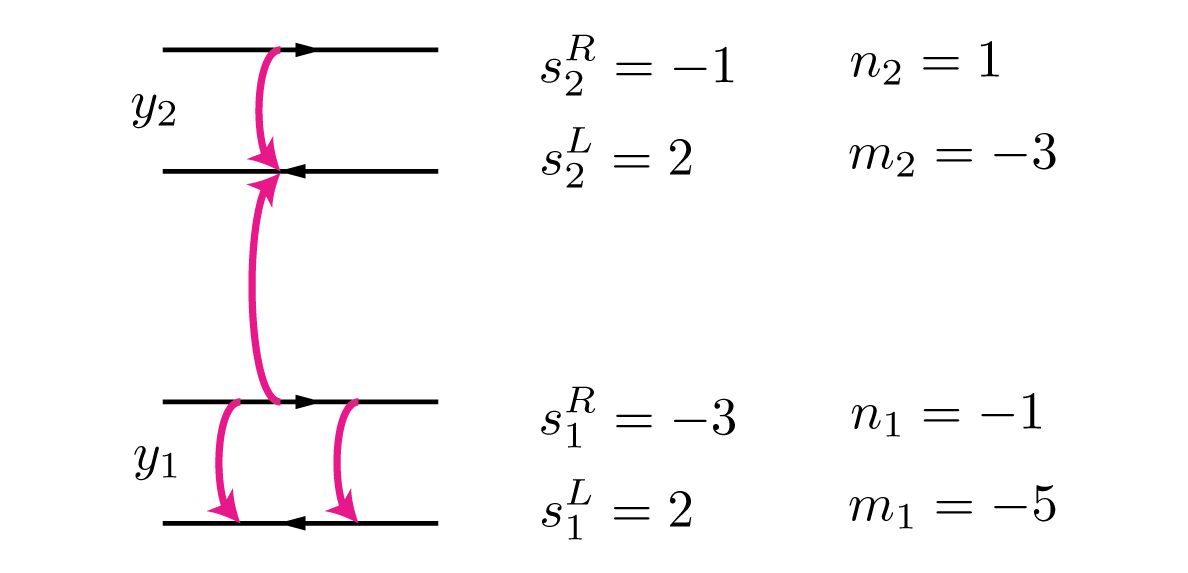}
    \caption{
    A diagram for an example coupling between two wires, showing left and right movers explicitly. Vertical arrows indicate the transfer of fermions between wires, and the description of the hopping is shown in terms of $s_j$, and equivalently in terms of  $n/m_j$. The $s_j$ parameters correspond exactly to the number of left and right fermions arriving on each wire, whereas $n_j$ denotes the total number of fermions arriving, and $m_j$ denotes the change of occupation from left to right on each wire. 
    }
        \label{fig:couplings_arrows}
\end{figure}

The couplings in \cref{eqn:coupling_operator} may be represented pictorially as shown in \cref{fig:couplings_arrows}a. It will be convenient to re-express these in terms of two new variables,
\begin{align}
    n_j &= s_j^{R} + s_j^{L},\label{eqn:n_def} \\
    m_j & = s_j^{R} - s_j^{L},\label{eqn:m_def}
\end{align}
where $n_j$ is the total number of fermions arriving at or leaving the wire, and $m_j$ encodes the transfer of fermions from $L$ to $R$ on a given wire. Note that $n_j$ and $m_j$ can take any value so long as both are even or both are odd,
\begin{align} \label{eqn:both_even_odd}
    m_j = n_j \mod 2.
\end{align}
Such couplings must generally satisfy charge and momentum conservation. To conserve charge, couplings must not produce a net change in the overall number of fermions,
\begin{align}\label{eqn:charge_conservation}
    \sum_{j} n_j = 0,
\end{align}
whereas to conserve momentum we require that
\begin{align} \label{eqn:momentum_conservation}
    \sum_j B y_j n_j + k_F m_j = 0.
\end{align}
We have freedom to choose the position of each wire, $y_j$. For a desired coupling between a set of $a$ wires we have two constraints and $a$ degrees of freedom. Thus, provided the coupling satisfies charge conservation, it is always possible to choose values of $\{y_j\}$ such that momentum is conserved. Every coupling thus implies a set of positions (or family of positions for $a>2$) for which this coupling conserves momentum. Thus, constructing an irregular array of wires necessitates irregular, spatially dependent couplings between wires.

\subsection{Bosonisation}

So far, we have constructed a generic Hamiltonian, represented by a set of linearised dispersive terms, $H^{\textup{Lin}}_j$, \cref{eqn:linearised_ham}, and some set of couplings $\mathcal O_{\{s_j\}}(x)$, Eq.~\eqref{eqn:coupling_operator}. The next step is to write down the bosonic reformulation of this system \cite{haldane_luttinger_1981,senechal_introduction_1999,von_delft_bosonization_1998}. We primarily use the conventions of Ref.~\cite{von_delft_bosonization_1998}, introducing a set of bosonic fields for left and right movers on each wire, $\phi_j^{L/R}$, which satisfy the following commutation relations
\begin{align}\label{eqn:bare_boson_commutation}
    [ \partial_x \phi_j^{p}(x) ,  \phi_{k}^{q}(x')] 
    = (-1)^p 2\pi i\, \delta_{jk} \delta_{pq} \delta(x-x').
\end{align}
We make use of the bosonisation identity to express our system in terms of these operators,
\begin{align}\label{eqn:bosonisation_identity}
    \psi^p_{j}(x) &\propto e^{-i\phi_j^p(x)},
\end{align}
where we have neglected to include the terms proportional to number density, as well as Klein factors \cite{von_delft_bosonization_1998,kane_fractional_2002,teo_luttinger_2014}. Furthermore, we use the following two identities for the kinetic term in $H^{\textup{Lin}}_j$ and the fermionic number density,
\begin{align}
    {\psi^p_j}^\dag(x) [-i\partial_x] \psi^p_j(x)
    &= (-1)^p \frac 12 {(\partial_x \phi_{j}^{p}(x))^2}, \\
    \rho^p_j= {\psi^p_j}^\dag(x)\psi^p_j(x)
    &= (-1)^p \partial_x \phi_j^{p}. \label{eqn:density_bosonisation}
\end{align}
Thus, the linearised Hamiltonian takes the following bosonic form,
\begin{align}\label{eqn:linearised_ham_bosonic}
    H_j^{\textup{Lin}} = \int \frac{\mathrm{d} x}{2\pi} 
    \frac v2 \left [
        \left (\partial_x \phi_{j}^{L}(x) \right ) ^2 +
        \left (\partial_x \phi_{j}^{R}(x) \right ) ^2
    \right ].
\end{align}
Finally, we can write an arbitrary coupling between wires in terms of the bosonic fields, by combining \cref{eqn:coupling_operator,eqn:bosonisation_identity}
\begin{align} \label{eqn:coupling_boson_fields}
    \mathcal O_{\{s_j\}}(x) \propto \cos\bigg (\sum_{j} s_j^L\phi_j^L(x)  + s_j^R\phi_j^R(x) \bigg ).
\end{align}

\subsubsection{Sum and Difference Fields}

Let us define a pair of bosonic sum and difference fields, which will often be more convenient than working in the $\phi^a_j$ basis, 
\begin{align}
    \varphi_j &= \frac 12 \left (\phi_j^R(x)+\phi_j^L(x)\right ), \label{eqn:Def_Sum_Field}\\
    \theta_j &= \frac 12\left (\phi_j^R(x)-\phi_j^L(x)\right ) \label{eqn:Def_Difference_Field},
\end{align}
which satisfy the commutation relations
\begin{align}
    \left [ \partial_{x} \theta_j(x) , \varphi_k(x') \right ] &=\pi i \delta_{jk} \delta(x-x'), \label{eqn:theta_comm_1}\\
    \left [ \partial_{x} \theta_j(x) , \theta_k(x') \right ] 
    &= \left [ \partial_{x} \varphi_j(x) , \varphi_k(x') \right ] 
    = 0.\label{eqn:theta_comm_2}
\end{align}
In terms of these fields, the linearised Hamiltonian, \cref{eqn:linearised_ham_bosonic}, takes the form
\begin{align}
    H_j^{\textup{Lin}} = \int \frac{\mathrm{d} x}{2\pi} 
    v \left [
        \left (\partial_x \varphi_{j}(x) \right ) ^2 +
        \left (\partial_x \theta_{j}(x) \right ) ^2
    \right ],
\end{align}
and each coupling, given by \cref{eqn:coupling_boson_fields}, may be written in terms of the $n_j$ and $m_j$ variables defined in \cref{eqn:n_def,eqn:m_def},
\begin{align}
\label{eqn:n_m_operators}
    \mathcal O_{\{s_j\}}(x) \propto \cos\bigg (
        \sum_{j}
        n_j \varphi_{j}(x) + m_j \theta_{j}(x)
    \bigg ).
\end{align}

Let us also consider the total density operator on a given wire,
\begin{align}
    \rho_j(x) = L^\dag_{j}(x)L_{j}(x)
    + R^\dag_{j}(x)R_{j}(x).
\end{align}
Written in terms of bosonic fields, using \cref{eqn:density_bosonisation}, this takes the form
\begin{align}\label{eqn:charge_density}
    \begin{aligned} 
    \rho_j(x) &= \partial_x \phi^R(x) - \partial_x \phi^L(x), \\
    &= 2 \partial_x \theta_j(x).
\end{aligned}
\end{align}
Thus, we see that the variation of $\theta_j(x)$ encodes the fermionic density on the $j$\ts{th} wire.

\section{Fractionalisation} \label{sec:fractionalisation}

In this section, we shall show how the construction above allows us to produce fractionalised fermionic states. First, we will show in \cref{sec:single_wire_frac} how the left and right-moving fermions on a single wire can be re-expressed in terms of a pair of fractionalised quasiparticle fields. By introducing an appropriate coupling between such fields, we then see how one can stabilise a fractionalised phase, where these quasiparticles make up the low-energy excitations. This discussion largely follows the standard method of Refs.~\cite{kane_fractional_2002,teo_luttinger_2014,tam_coupled_2020,tam_nondiagonal_2021}.
Then in \cref{sec:gluing_fractions} we show how it is possible to construct composite materials by connecting wires with different fractionalisation. Crucially, we will see that it is possible to introduce couplings that can gap out left edge modes from one wire with right edge modes from another, where the two wires have different fractionalisation. 

\subsection{Uniform Fractionalised Phases}\label{sec:single_wire_frac}

\begin{figure}[t]
    \centering
    \includegraphics{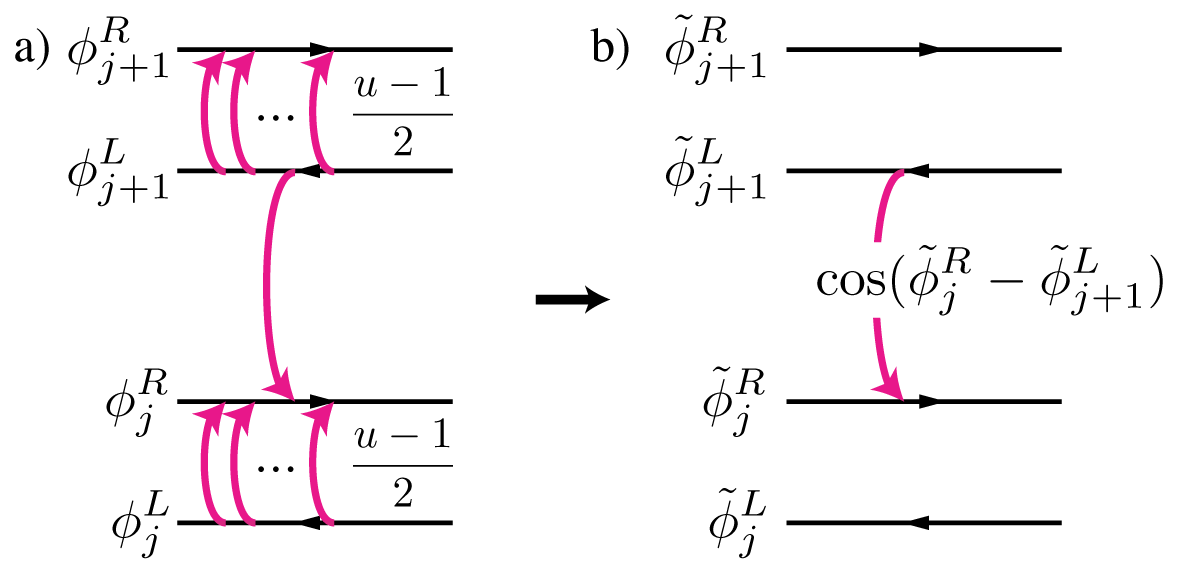}
    \caption{
        (a) The coupling that stabilises a $1/u$ Laughlin phase.
        (b) By locally rotating the Hilbert space on each wire in terms of new $\tilde \phi^{R/L}$ fields, the coupling from subplot (a) is re-expressed as a simple locking term between two fractionalised fields, see Eq.~\eqref{eqn:tilde_coupling}. 
    }
    \label{fig:couplings_laughlin}
\end{figure}

We start by introducing a change of basis on each wire, from our \textit{bare} $\theta$ and $\varphi$ fields to a new set of fractionalised bosonic fields. This transformation is defined by a parameter $u \in \mathbb N$. 
\begin{align} 
    \tilde \phi^R_j(x) &= \varphi_{j}(x) +u\theta_{j}(x), \label{eqn:phi_R_tilde} \\
    \tilde \phi^L_j(x) &= \varphi_{j}(x) -u\theta_{j}(x).\label{eqn:phi_L_tilde}
\end{align}
These new fields satisfy similar commutation relations to the bare fields, \cref{eqn:bare_boson_commutation}, although they are rescaled by a factor of $u$,
\begin{align} \label{eqn:fractionalised_commutators}
    [\partial_x \tilde \phi^{p}_j(x), \tilde \phi^{q}_k(x')] = \mp 2\pi i u \delta_{jk} \delta_{pq}\delta(x-x').
\end{align}
As we shall see in \cref{sec:general_construction}, these fields correspond to a pair of $1/u$-fractionalised chiral left and right movers. 

Now all that remains is to construct an interaction that couples these fractionalised right and left movers between adjacent wires, which will ensure that the system as a whole is in a $1/u$ fractionalised phase. We consider a coupling between the wires at $j$ and $j+1$ given by the parameters
\begin{align}
    m_j = m_{j+1} &= u,\\
    n_j &= +1,\\ 
    n_{j+1} &= -1.
\end{align}
A diagram of this coupling is shown in \cref{fig:couplings_laughlin}a.
In terms of bosonic fields, this leads to a term of the form
\begin{align}
    \mathcal O(x) = \cos\left (
        \varphi_{j} 
        +u\theta_{j} 
        - \varphi_{j+1} 
        + u\theta_{j+1}  
        \right ).
\end{align}
Comparing to \cref{eqn:phi_R_tilde,eqn:phi_L_tilde}, we see that this coupling can be written as
\begin{align} \label{eqn:tilde_coupling}
    \mathcal O(x) = \cos\left (
        \tilde \phi^R_j
        - \tilde \phi^L_{j+1} 
        \right ),
\end{align}
where we have coupled a fractionalised right mover from one wire onto the left mover on the wire above. Next, we transform to a new set of sum and difference fields,
\begin{align}
    \tilde \varphi_{j+\frac 12 } &= \frac 12 \left (\tilde \phi_j^R(x)+\tilde \phi_{j+1}^L(x)\right ),\label{eqn:uniform_varphi_tilde}\\
    \tilde \theta_{j + \frac 12 } &= \frac 12\left (\tilde \phi_j^R(x)-\tilde \phi_{j+1}^L(x)\right ).\label{eqn:uniform_theta_tilde}
\end{align}
These sum and difference fields exist at the boundary between the $j$\ts{th} and $(j+1)$\ts{th} wires, which we will define as the $j+\frac 12$ \textit{quasiwire}. In what follows, bare fermion $\theta$ and $\phi$ fields will be defined on the original wires, whereas the fractionalised $\tilde \theta$ and $\tilde \phi$ fields exist on these quasiwires.

We find that these have almost identical commutation relations to those satisfied by $\theta$ and $\varphi$, see \cref{eqn:theta_comm_1,eqn:theta_comm_2}, aside from an additional factor of $u$, 
\begin{align}
     [ 
        \partial_{x} \tilde \theta_{j + \frac 12 }(x) , 
        \tilde \varphi_{k + \frac 12 }(x') 
     ] &=
    \pi i u \delta_{jk} \delta(x-x'), \label{eqn:theta_phi_commutator}\\
     [
         \partial_{x} \tilde \theta_{j + \frac 12 }(x) , 
         \tilde \theta_{k + \frac 12 }(x') 
     ] 
    &=0 ,\label{eqn:theta_self_commutator} \\
     [
         \partial_{x} \tilde \varphi_{j + \frac 12 }(x) , 
         \tilde \varphi_{k + \frac 12 }(x') 
     ] 
    &= 0.\label{eqn:phi_self_commutator}
\end{align}
Thus, the coupling becomes
\begin{align}
    \mathcal O(x) = \cos\left (
        2\tilde \theta_{j + \frac 12 }
        \right ).
\end{align}
This operator satisfies two properties that are necessary for the system to admit an exact solution: $\tilde \theta_{j + \frac 12 }$ must commute between wires, and $\tilde \theta_{j + \frac 12 }(x)$ must commute with $\tilde \theta_{j + \frac 12 }(x')$ when $x \neq x'$. This ensures that all couplings across the $x$ and $y$ directions commute with one another, and thus can be independently satisfied in the ground state.

\subsection{Gluing Fractions Together}\label{sec:gluing_fractions}

\begin{figure}[t]
    \centering
    \includegraphics{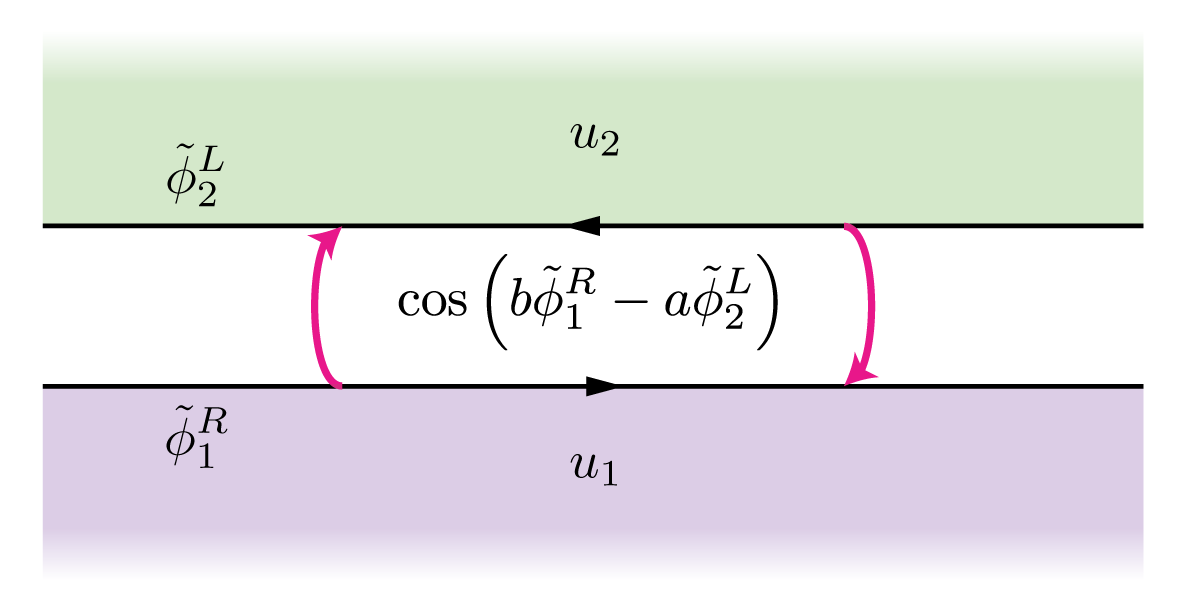}
    \caption{
        Two regions of coupled wires where the lower (purple) region forms a $1/u_1$ fractionalised Laughlin phase and the upper (green) region forms a $1/u_2$ phase. A coupling is introduced between the two respective edge modes that pairs $b$ $u_1$-quasiparticles with $a$ $u_2$-quasiparticles. By choosing appropriate values of $a$ and $b$ such that $u_1/u_2 = a^2 / b^2$, see Eq.~\eqref{eq:ucond}, we can ensure the boundary remains gapped.  
    }
    \label{fig:gluing_fractions}
\end{figure}

Now, let us consider an interface between two sets of coupled wires, where we wish to enforce different fractionalisations ($u_1$ and $u_2$) in each wire, as shown in \cref{fig:gluing_fractions}. Thus, we wish to couple a $u_1$-fractionalised mode, $\tilde \phi^R_1$, from the lower wire with a  $u_2$-fractionalised mode, $\tilde \phi^L_{2}$, from the upper wire. This coupling must take an analogous form to \cref{eqn:tilde_coupling}, which preserves the exact solubility of the system. With that in mind, let us propose a trial coupling
\begin{align}\label{eqn:p_q_coupling}
    \mathcal O(x) = \cos\left ( 
        b \tilde \phi^R_1 -  a \tilde \phi^L_{2}
    \right ),
\end{align}
where $a$ and $b$ are a pair of parameters that we must choose. Note that we require $a,b \in \mathbb Z$ for it to be possible to express this coupling in terms of bare fermions.

This coupling determines a pair of operators that play the same role as the $\tilde \theta$ and $\tilde \varphi$ operators presented in \cref{eqn:uniform_varphi_tilde,eqn:uniform_theta_tilde},
\begin{align}
    \tilde\theta_{1.5} = \frac 12 (b \tilde \phi^R_1 -  a \tilde \phi^L_{2}),\\ 
    \tilde\varphi_{1.5} = \frac 12 (b \tilde \phi^R_1 +  a \tilde \phi^L_{2}).
\end{align}
These should obey an analogous set of commutation relations as \cref{eqn:theta_phi_commutator,eqn:theta_self_commutator,eqn:phi_self_commutator}. In particular, for the coupling to preserve exact solubility, we require that $\tilde \theta_{1.5}$ commute with itself at different positions in space. Calculating this explicitly for $\tilde \theta$, we find that
\begin{align}
    [\partial_x\tilde \theta_{1.5}(x),
    \tilde \theta_{1.5}(x')] = \frac{\pi i} 4 (u_1 b^2  - u_2a^2 ) \delta(x-x').
\end{align}
Thus, by setting
\begin{align}
\label{eq:ucond}
    \frac{u_1}{u_2} = \left (\frac ab \right )^2
\end{align}
we ensure that $\tilde \theta_{1.5}$ satisfies the desired commutation relations, and so can have a well-defined expectation value. Since all variables here must be $\in \mathbb Z$. Thus, it is only possible to introduce such a coupling between a pair of fractionalised phases if the ratio between fractions is a square of a rational number. This is consistent with the so-called \textit{null-vector criterion} proposed by Haldane \cite{haldane_stability_1995}, which is also discussed in detail in \cite{santos_parafermionic_2017,may-mann_families_2019}. Following this condition, let us write the fractionalisation coefficients $u_1$ and $u_2$ in terms of a common factor $u$, and our pair of integers $a$ and $b$, according to
\begin{align}
    u_1 &= ua^2, \\
    u_2&= ub^2,
\end{align}
We may now write the full commutation relations of the quasiparticle sum and difference operators, analogous to \cref{eqn:theta_phi_commutator,eqn:theta_self_commutator,eqn:phi_self_commutator},
\begin{align}
    [ \partial_{x} \tilde \theta_{1.5}(x) , 
    \tilde \varphi_{1.5}(x') ] &=
    \pi i u a^2b^2  \delta(x-x'),\\
    [\partial_{x} \tilde \theta_{1.5}(x) , 
        \tilde \theta_{1.5}(x') ] 
    &= 
    [\partial_{x} \tilde \varphi_{1.5}(x) , 
    \tilde \varphi_{1.5}(x') ] 
    = 0,
\end{align}
and the coupling between the two regions can be written in a familiar form
\begin{align}
    \mathcal O(x) = \cos\left (
        2 \tilde \theta_{1.5}(x)
    \right ).
\end{align}

\subsubsection{Charge Conservation} \label{sec:charge_conservation}

Remarkably, we have seen that it is possible to introduce a coupling between two fractionalised phases with different fractionalisation, such that the boundary between the phases is gapped. However, such a coupling comes with a price, which is that we have violated charge conservation. To see how, let us convert our coupling back into \textit{bare fermion} sum and difference fields using the identities
\begin{align}
    \tilde \phi_1^R &= \varphi_1 + ua^2 \theta_1, \\
    \tilde \phi_2^L &= \varphi_2 - ub^2 \theta_2,
\end{align} 
such that the coupling becomes
\begin{align}
\label{eq:Interface_Coupling}
    \mathcal O(x) = \cos\left (
        b \varphi_1 - a\varphi_2
        +ua^2b\theta_1 + uab^2\theta_2
    \right ).
\end{align}

Thus, following the discussion in \cref{sec:allowed_couplings}, we see that such a term involves a net addition (or subtraction in the case of the hermitian conjugate) of $(a-b)$ bare fermions to the system. Taken alone, this coupling does not respect charge conservation~\cite{santos_parafermionic_2017, may-mann_families_2019}. Two solutions may be proposed to this problem. 
Firstly, as is shown in \cref{apx:substrate}, charge conservation may be restored by coupling to a superconducting substrate to allow for the exchange of cooper pairs between system and substrate. This may be modelled by including an additional non-fractionalised wire which exchanges charge with our system, but changes none of the properties of the fractionalised phase. Thus, we neglect to include it explicitly in the following analysis. 
Otherwise, we may also consider, as is discussed in \cite{santos_parafermionic_2017}, starting with a composite particle which already carries a multiple of an electron's charge, such that the charge conservation is restored under the couplings shown here. We shall not consider this possibility any further and leave it to future work.

\subsubsection{Momentum Conservation}

We may additionally want to check that this coupling satisfies momentum conservation. Comparing with \cref{eqn:momentum_conservation}, we see that the condition imposed is
\begin{align}
\label{eq:momconservmain}
    B(b y_1 - a y_2) + k_Fuab(a + b) = 0.
\end{align}
Since we are free to choose the values of $y_j$, this condition is always satisfiable. Furthermore, this relation provides a correspondence between the relative fractionalisation of each wires and their positions in real space. Thus, in order to produce a composite phase where different wires have different fractionalisation, we necessarily must place wires at non-uniform separation from one another.

\section{A General Construction for Non-Uniform Fractionalisation}\label{sec:general_construction}

We now have all the ingredients necessary to construct a generic composite fractionalised system. That is, given a set of wires, we have shown that it is possible to re-express the degrees of freedom of each wire in a $ 1/{u_j}$ fractionalised basis \footnote{Of course, for fermionic systems we require $u_j$ odd, however the same can be done starting with bosons to obtain even fractions.}. Additionally, we have seen that, provided all such fractions are related by a square, taking the form $u_j = ua_j^2$, with a common $u$ and all $a_j$ coprime to their adjacent wires, one can to introduce a coupling between each pair of wires such that the system forms an overall gapped phase. This construction involves the series of transformations given above, which we compile in \cref{apx:glossary_of_identities} for ease of use. 

We now study the properties of excitations that arise within such a construction, deriving the `charge' carried by these particles, the constraints on their transport and their mutual statistics. 
In \cref{sec:quasiparticle_charge} we review the definition of charge carried by every quasiparticle in the uniform Laughlin phase, following \cite{kane_fractional_2002}. Next, extending to the disordered context, a notion of overall charge cannot be defined (see \cref{sec:charge_conservation}), however we define and calculate a quantity analogous to charge, which is conserved.
Finally, in \cref{sec:composite_anyons} we discuss the braiding of excitations in such a composite fractional material. We explicitly calculate their exchange statistics and show how the system divides into composite fractionalised anyons which are free to move throughout the system, and lineons which can only propagate in the $x$-direction. 

Let us start by establishing the most generic configuration, in which we have an array of $N$ wires at positions $y_j$. The couplings to each wire are bind some multiple of a $1/{u{a_j^2}}$-fractionalised set of left and right movers. Thus, we write the degrees of freedom of each wire in terms of left and right-moving fractionalised bosonic fields, $\tilde \phi_{j}^{L/R}$, whose commutator includes a factor of $ua_j^2$, following \cref{eqn:fractionalised_commutators}.

We place a coupling between each pair of adjacent wires that takes the form $\cos(2\tilde \theta_{j + \frac 12 })$, with
\begin{align}
    \tilde \theta_{j + \frac 12 } & = \frac 12  \left (
        a_{j+1} \tilde \phi_{j}^R 
        - a_{j}\tilde \phi_{j+1}^L
    \right ), \label{eqn:compound_theta_tilde}\\ 
    \tilde \varphi_{j + \frac 12 } & = \frac 12  \left (
        a_{j+1} \tilde \phi_{j}^R 
        + a_{j}\tilde \phi_{j+1}^L
    \right ).\label{eqn:compound_varphi_tilde}
\end{align}
Thus, the overall Hamiltonian is
\begin{align} \label{eqn:couplings_def}
    H = H_{\textup{Kin}} + \int \frac {dx}{2\pi}
    \sum_{j} g\cos \left (
        2 \tilde \theta_{j + \frac 12 }(x)
    \right ),
\end{align}
where $H_{\textup{Kin}}$ is a quadratic function in $\{\partial_x\tilde \theta_{j + \frac 12 }\}$, and $\{\partial_x\tilde \varphi_{j + \frac 12 }\}$, and $g$ is a constant that determines the coupling strength. Let us parametrise $H_{\textup{Kin}}$ in terms of a $2N \times 2N$ matrix $M$ as
\begin{align}
\label{eqn:Main_Hkin}
    H_{\textup{Kin}} = \int \frac {dx}{2\pi} 
    \begin{pmatrix}
        \partial_x \tilde {\bm \varphi} \\
        \partial_x \tilde {\bm \theta} 
    \end{pmatrix}^T
    \begin{pmatrix}
        M^{\varphi\varphi} & M^{\theta\varphi} \\
        M^{\varphi\theta} & M^{\theta\theta} 
    \end{pmatrix}
    \begin{pmatrix}
        \partial_x \tilde {\bm \varphi} \\
        \partial_x \tilde {\bm \theta} 
    \end{pmatrix},
\end{align}
where $\tilde{\bm \varphi}$ represents the column vector $\begin{pmatrix}
    \tilde \varphi_{0.5}  &...&\tilde\varphi_{N-\frac 12}
\end{pmatrix}^T$, with a similar expression for $\tilde {\bm \theta}$.

Since the coupling is expressed exclusively in terms of the difference field $\tilde \theta$, the only condition for our solution to remain valid is that terms involving $\tilde \varphi$ (which does not commute with $\tilde \theta$) are weak. In the limit $M^{\varphi\varphi} = M^{\theta\varphi} = M^{\varphi\theta}=0$, all terms in $H$ commute precisely, and we may treat $\tilde \theta$ as a classical variable, replacing it with its expectation value. Consequently, the system sits in a gapped ground state defined by the lowest energy configuration of these $\{\tilde \theta_{j+\frac 12}\}$ fields.

Provided that no terms in $H_{\textup{Kin}}$ involving $\tilde \varphi$ are strong enough to close the gap to the $\tilde \theta$-ordered ground state, we can effectively treat the system as fluctuating close to this exact, effectively classical ground state. In what follows, we shall assume that we are working in this limit, the conventional limit where such bosonisation constructions are valid \cite{kane_fractional_2002,teo_luttinger_2014,tam_nondiagonal_2021}. Furthermore, we must also ensure that our couplings, \cref{eqn:couplings_def}, remain relevant under renormalisation group (RG) flow. This depends strongly on the form of $H_{\textup{kin}}$, so we must ensure that it is possible to construct a kinetic term such that the desired couplings are the most relevant possible term that satisfies momentum conservation. This is provided in \cref{apx:renormalisation}.

At this point, the Hamiltonian has been reduced to an interplay between two sets of terms: coupling terms, which are minimised when all fields are fixed to a multiple of $\pi$, and kinetic terms quadratic in $\partial_x \tilde \theta$, which penalise spatially changing $\tilde \theta$ and vanish when it is constant. Thus, it is straightforward to write down the ground state which satisfies both terms: All fields are spatially uniform and pinned to a multiple of $\pi$,
\begin{align}
\label{eq:pinning}
    \tilde \theta_{j + \frac 12 }(x) = \tilde \theta_{j + \frac 12 } \in \pi \mathbb Z.
\end{align}

\begin{figure*}[ht!]
    \centering
    \includegraphics{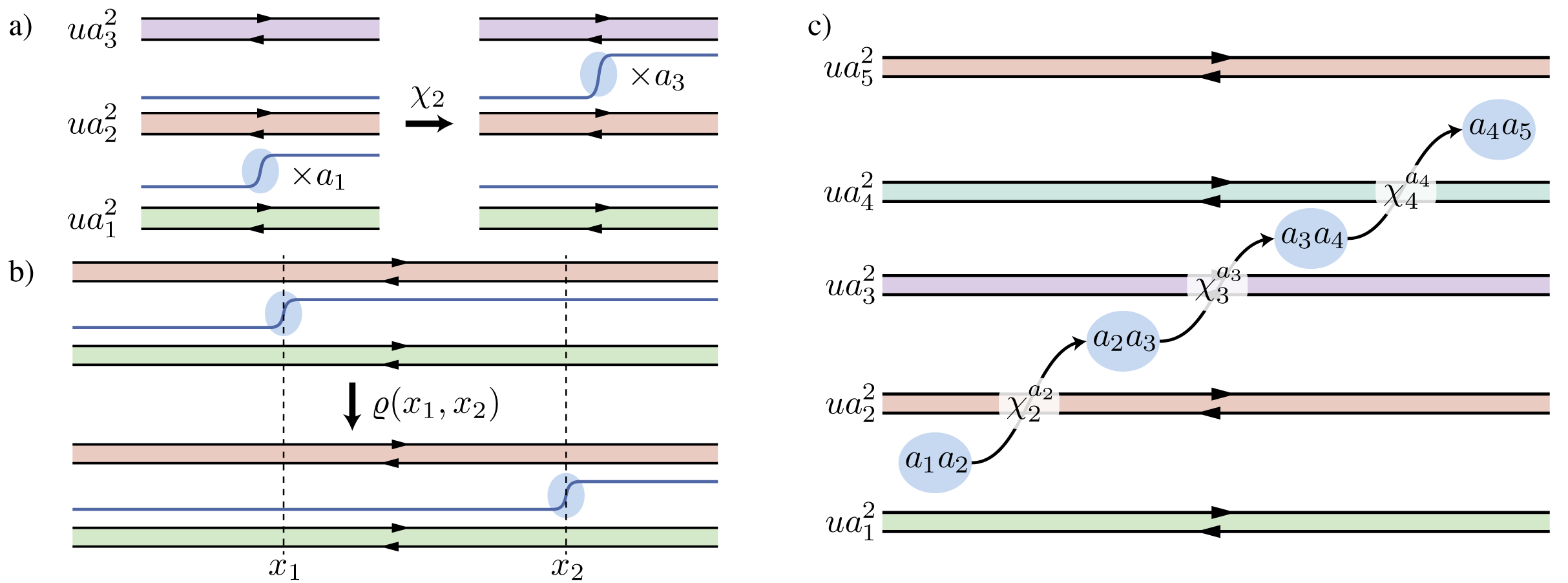}
    \caption{
        Transport of anyons through a system of fractionalised wires. Each wire may be differently fractionalised with $ua_j$, indicated by colour, as well as being unevenly displaced from one another.
        (a) The $\chi$ operator,        \cref{eqn:chi_as_right_left}, is responsible for transport in the $y$-direction. On a set of three coupled wires, the $\chi_2 (x)$ operator destroys a composite of $a_1$ quasiparticles on the $\tilde \theta_{1.5}$ quasiwire, and creates a composite of $a_3$ quasiparticles from the $\tilde \theta_{2.5}$ quasiwire.
        (b) The $\rho$ operator, Eq.~\eqref{eqn:Rho_operator}, is responsible for transport in the $x$-direction, where the operator $\rho(x_1,x_2)$ destroys a single kink at position $x_1$ and creates one at $x_2$ on the same wire.
        (c) On a given $\tilde\theta_{j + \frac 12 }$ quasiwire, a composite excitation with $a_ja_{j+1}$ multiplicity can travel freely to the quasiwire above and below by applying the $\chi_j$ operator $a_j$ times. Furthermore, the nature of the $\chi$ operator is such that it ensures the new quasiparticle is always created with the correct multiplicity to allow it to travel onto the wire beyond. 
    }
    \label{fig:chi_and_rho}
\end{figure*}

\subsection{Quasiparticle Charge}\label{sec:quasiparticle_charge}

Quasiparticles in the bosonised system correspond to a kink in $\tilde \theta_{j + \frac 12 }$, where the value jumps by $\pm\pi$. 
As a warm-up, let us start by determining the bare charge carried by quasiparticles in the well-studied uniform case (all $a_j = 1$), and then move on to the more involved non-uniform case.

{\it The Uniform Case:} Following \cref{eqn:charge_density}, we may write down the expression for the total \textit{bare fermion} charge contained on the $j^{\textup{th}}$ wire within a region $x_I \rightarrow x_F$ as
\begin{align}
\label{eq:totalcharge}
    Q_j(x_I, x_F) &= 
    \int_{x_I}^{x_F}\frac{dx}{2\pi} \rho_{j}(x), \\ 
    &=   \left [ \frac {\theta_j (x) }\pi \right ]_{x_I}^{x_F}.
\end{align}
Using \cref{eqn:phi_L_tilde,eqn:phi_R_tilde,eqn:uniform_varphi_tilde,eqn:uniform_theta_tilde} we may now re-express the bare density in terms of our fractionalised density variables $\tilde \varphi$ and $\tilde \theta$,
\begin{align}
    \frac {\theta_j (x) }\pi = \frac{1}{2u\pi } 
    \left (
        \tilde \varphi_{j + \frac 12 } 
        - \tilde \varphi_{j-\frac 12} 
        + \tilde \theta_{j + \frac 12 }
        + \tilde \theta_{j-\frac 12}
    \right ).
\end{align}
Now, taking a sum over a number of wires from $j = 1$ to $j = N$, we find that terms with $\tilde \varphi$ cancel out in the bulk of the region, leaving 
\begin{align}\begin{aligned}
    \sum_{j = 1}^N \frac {\theta_j (x) } \pi & =  
    \frac 1{u} \sum_{j = 1}^{N-1} \frac{\tilde \theta_{j + \frac 12 }} \pi 
    + \left ( 
        \textup{boundary terms}
    \right ),
\end{aligned}
\end{align}
where the boundary terms exist only on the $1^{\textup{st}}$ and $N^{\textup{th}}$ wires, so we neglect them. Thus, we see from Eq.~\eqref{eq:totalcharge} that a kink where $\tilde \theta_{j + \frac 12 }$ changes by $\pi$ within the bulk of the region must correspond to adding a fractional charge of $1/u$.

{\it The Non-uniform Case:} We may once again re-express the bare fermion density in terms of fractionalised fields, although we now must use \cref{eqn:compound_theta_tilde,eqn:compound_varphi_tilde}, to get
\begin{align}
    \frac{\theta_j}{\pi} = \frac{1}{2ua_j\pi} 
    \left (
        \frac{\tilde \varphi_{j + \frac 12 }} {a_j a_{j+1}} 
        - \frac{\tilde \varphi_{j-\frac 12}} {a_j a_{j-1}} 
        + \frac{ \tilde \theta_{j + \frac 12 }} {a_j a_{j+1}}
        + \frac{ \tilde \theta_{j-\frac 12}} {a_j a_{j-1}}
    \right ).
\end{align}
By considering the denominators in this expression, we see that there are no neat cancellations in $\sum_j{\theta_j}$, which now retains dependence both on the $\tilde \varphi$ and $\tilde \theta$ fields in the bulk. This is unsurprising, since we have already established that bare fermionic charge is not conserved in the composite system, see \cref{sec:charge_conservation}, unless a substrate is included, see \cref{apx:substrate}. Thus, we should not be surprised to see that overall charge does not have a well-defined expectation value. Rather, let us consider the quantity
\begin{align}
    \sum_{j = 1}^N \frac{a_j \theta_j}\pi & = 
    \frac 1{u\pi} \sum_{j = 1}^{N-1} 
    \frac {\tilde \theta_{j + \frac 12 }}{ a_j a_{j+1}}
    + 
    \left ( 
        \textup{boundary terms}
    \right ). \label{eqn:compound_conserved_charge}
\end{align}
Since this is expressed only in terms of $\tilde \theta$, it is conserved and has a well-defined expectation value close to the ground state.

\subsection{Composite Statistics}\label{sec:composite_anyons}

In this section, we calculate anyonic braiding statistics. This calculation takes three parts. In the first we establish the operators necessary to transport anyons throughout the system. Next, we see how the properties of these transport operators lead to the emergence of two types of excitation: a set of composite anyons that are free to travel throughout the system along both axes, and a set of lineons which are confined to move only in the $x$-direction~\cite{Nandkishore_2019,Pretko_2020,gromov_fractons_2024}. Finally, we braid both sets of anyons and calculate relative statistics.

\subsubsection{Moving Anyons}\label{sec:moving_anyons}

{\it Between Quasiwires---}%
Let us first consider the backscattering of a single fermion from $R$ to $L$ on a single wire, represented by the following operator,
\begin{align}\begin{aligned}
    \chi_j(x) =  &{\psi^L_j}^\dag(x) {\psi^R_j}^{\phantom{}}(x) \\
    &\rightarrow e^{-i[\phi^R_j(x) - \phi^L_j(x)]}.
\end{aligned}
\label{eqn:Chi_operator}
\end{align}
Re-expressed in terms of fractionalised fields, this takes the form
\begin{align} \label{eqn:chi_as_right_left}
    \chi_j(x) = \exp\left(\frac {-i}{ua_j^2}
    [\tilde \phi^R_j(x) - \tilde \phi^L_j(x)]\right).
\end{align}
We now compute the effect of $\chi_j (x) $ on the $\partial_x \tilde \theta (x')$ fields,
\begin{align}
    [\partial_{x} \tilde \theta_{j + \frac 12 }(x),
    \chi_j(x')]
    &= \pi a_{j+1}\delta(x-x')\chi_j(x') , \label{eqn:chi_above}
    \\ 
    [\partial_{x} \tilde \theta_{j-\frac 12}(x),
    \chi_j(x')]
    &= -\pi a_{j-1}\delta(x-x')\chi_j(x'),\label{eqn:chi_below}
\end{align} 
which we use to derive the following identities,
\begin{align}
    \chi_j^{-1}\partial_{x} \tilde \theta_{j + \frac 12 }
    \chi_j
    = \partial_{x} \tilde \theta_{j + \frac 12 }
    +\pi a_{j+1}\delta(x-x'), \\
    \chi_j^{-1}\partial_{x} \tilde \theta_{j - \frac 12 }
    \chi_j
    = \partial_{x} \tilde \theta_{j - \frac 12 }
    -\pi a_{j-1}\delta(x-x'), 
\end{align}
where $\chi$ acts at $x'$ and $\partial \tilde \theta$ acts at $x$. Thus, we see see that the operator removes $a_{j-1}$ quasiparticles on the $\tilde\theta_{j-\frac 12}$ quasiwire and creates $a_{j+1}$ quasiparticles on the $\tilde\theta_{j + \frac 12 }$ quasiwire. A diagram is shown in \cref{fig:chi_and_rho}a. Thus, we see that in the $y$-direction, quasiparticles can only be moved in groups of $a_j$.

{\it Along Quasiwires---}%
The operator that transfers
a quasiparticle from $x_1$ to $x_2$ along the $j+1/2$ quasiwire is
\begin{align}
\label{eqn:Rho_operator}
    \varrho_{j+\frac 12}(x_1, x_2) &= \exp\left(i\frac{
        \tilde \varphi_{j + \frac 12 }(x_1) 
    -
    \tilde \varphi_{j + \frac 12 }(x_2) 
    }{{ua_{j}^2 a_{j+1}^2}}\right),\\
    &= \exp\left(i\frac{\int_{x_2}^{x_1}\mathrm{d}x\, \partial_x\tilde{\varphi}_{j+1/2}(x)}{ua_j^2 a_{j+1}^2}\right),
\end{align}
which is a local operator since $\partial_x\tilde{\varphi}$ can always be written in terms of the bare fermion densities, see \cref{eqn:density_bosonisation}.
This operator allows us to transport a \textit{single} $\pm \pi$ kink along a wire to any arbitrary position. 

\subsubsection{Lineons and Composite Anyons} \label{sec:lineons_and_composite}
\begin{figure}[t]
    \centering
    \includegraphics{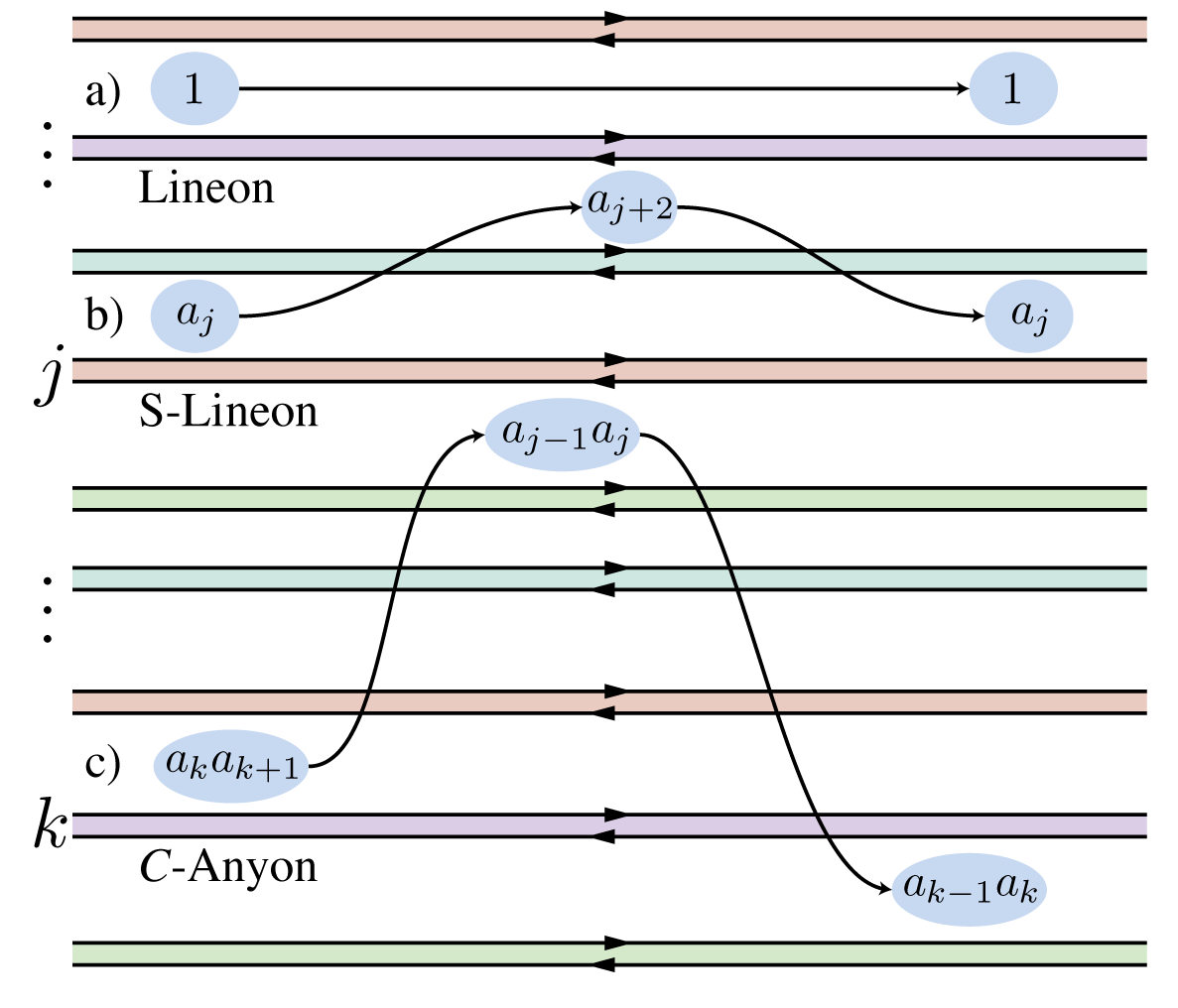}
    \caption{
        Three types of fractionalised excitations in a system of unevenly spaced wires, where the $j$\ts{th} wire is in a $1/ua_j^2$ phase.
        (a) A single kink between the $j$ and $j+1$ wire corresponds to a quasiparticle excitation that is constrained to only move in the $x$ direction.
        (b) A composite of $a_j$ kinks between the $j$ and $j+1$ wire is only able to travel up to the next wire and back. 
        (c) Finally, a $C$-anyon, a composite of $a_ka_{k+1}$ kinks has complete freedom to travel throughout the system. Note that (a) and (b) are fractons, constrained to move in the $x$-direction, whereas (c) is not constrained. 
    }
    \label{fig:anyon_types}
\end{figure}

Following the form of the above quasiparticle transport operators \cref{eqn:chi_as_right_left,eqn:Rho_operator}, we see that the $x$ and $y$-directions in our model effectively `see' a different type of quasiparticle. In the $x$-direction, each quasiparticle is free to move independently along the wire. However, in the $y$-direction quasiparticles can only be transported together in groups of $a_j$, where each interface between wires with a different value of $a_j$ thus transfers a different number of quasiparticles, as shown in \cref{fig:chi_and_rho}. Thus, it is natural to ask if there is a minimal combined quasiparticle that is free to travel throughout the system in both directions. 

To answer this, we consider a set of $\eta$ quasiparticles at the same position on a quasiwire at $\tilde \theta_{j + \frac 12 }$. In order to be able to travel freely in the $y$-direction, we must at least be able to hop this set of quasiparticles both onto the quasiwire above and the one below. Thus, there are two $\chi$ operators to consider, following \cref{eqn:chi_above,eqn:chi_below}. To move the bundle up to the next wire, we must apply a $\chi_{j+1}$ operator, which destroys $a_j$ kinks and creates $a_{j+2}$ kinks on the next quasiwire. Similarly, to move the bundle down, we must apply a $\chi_{j}^\dag$ operator, which destroys $a_{j+1}$ kinks and creates $a_{j-1}$ on the quasiwire below. Thus, for this bundle of quasiparticles to be freely mobile, we must have started with a number $\eta$ of particles in the bundle that is both divisible by $a_j$ and $a_{j+1}$. Since all $a_j$ are coprime, the minimal acceptable bundle contains $a_j\times a_{j+1}$ quasiparticles. 

Now, suppose we start with such a bundle of $a_j\times a_{j+1}$ quasiparticles on the $\tilde \theta_{j + \frac 12 }$ quasiwire and wish to transport it to the quasiwire above. Since $\chi_{j+1}$ only moves $a_j$ individual quasiparticles, we will have to apply it $a_{j+1}$ times. Each time we destroy $a_j$ particles and create $a_{j+2}$ particles on the next wire. Thus, we finish having removed the full bundle, and created a bundle of $a_{j+1}\times a_{j+2}$ quasiparticles on the quasiwire above, at $\tilde \theta_{j+1,j+2}$. Remarkably, the transport operators conspire to ensure that the bundle created on the next wire has the right multiplicity to allow it to hop onto the wire above, and so on. A diagram of this process is shown in \cref{fig:chi_and_rho}c.

    Excitations in the system can thus be classified broadly into three types. One has a set of fractionalised quasiparticles corresponding to a single $\pm \pi$ kink on a given quasiwire. These are constrained to only move in the $x$-direction -- constituting a set of fractonic excitations called lineons~\cite{Nandkishore_2019,Pretko_2020}. Next, we may combine a group of $a_j$ such excitations to form a constrained compound particle, which can only hop over one adjacent wire, but no further. These have also constrained motion as lineons, but are spread among two wires, and hence we refer to them as spread-lineons, or s-lineons for short. Finally, combining $a_j\times a_{j+1}$ fractonic quasiparticles on the $\tilde \theta_{j + \frac 12 }$ quasiwire produces a free composite quasiparticle that may travel unimpeded throughout the system. We shall label these as $C$-particles. These three cases are shown in \cref{fig:anyon_types}.

\subsubsection{Braiding}\label{sec:braiding}

Let us now calculate the braiding statistics of the anyons in our system. %
Only $C$-type composite particles can be transported freely around the system, whereas lineons can only participate by having a $C$-type quasiparticle moved around them.
Note that, for the braiding statistics to pick up only a topological phase, one must perform braiding with all quasiparticles well-separated in space. Bringing kinks close to one another changes the energy of the system, causing the phase to precess. To avoid this, we assume that quasiparticles are always well-separated throughout the braiding operation.

To determine the braiding statistics, we must first examine the $\chi$ operators responsible for moving a kink between wires. Starting with \cref{eqn:chi_as_right_left}, we rewrite the term in the exponential as
\begin{align}\label{eqn:phi_tilde_r_minus_l}
    \tilde \phi^R_j(x) - \tilde \phi^L_j(x) 
    = 
        \frac{\tilde \varphi_{j + \frac 12 }+
        \tilde \theta_{j + \frac 12 }}
        {a_{j+1}}
    -
        \frac{ \tilde \varphi_{j-\frac 12}-
        \tilde \theta_{j-\frac 12}}
        {a_{j-1}}.
\end{align}
Next, we consider the operation that moves a single $C$-particle throughout the system, shown in \cref{fig:chi_and_rho}c, where we apply each $\chi_j$ operator $a_j$ times.
Since $\chi$ on different wires commute, we may write the overall operator that transports a $C$-particle from the $(j-\frac 12)$\ts{th} to the $(k+\frac 12)$\ts{th} quasiwire as the following product of operators,
\begin{align}
    \prod_{i=j}^k \chi_{i}^{a_i}(x) = \exp\left ( 
        \frac {-i}{u}
        \sum_{i=j}^k    
        \frac{\tilde \phi^R_i(x) - \tilde \phi^L_i(x)}{a_i}
    \right ),
\end{align}
inserting \cref{eqn:phi_tilde_r_minus_l}, we find that all $\tilde \varphi$ operators cancel out aside from those at the endpoints,
\begin{align} \label{eqn:theta_sum_term}
    \prod_{i=j}^k \chi_{i}^{a_i}(x) = 
    \exp\left(
    -i
        \sum_{i=j}^{k-1}
        \frac{2 \tilde \theta_{i+\frac 12}}
        {u a_ia_{i+1}}
        + \hat{\mathcal O}_{\textup{endpoints}}
    \right),
\end{align}
with 
\begin{align}
    \hat{\mathcal O}_{\textup{endpoints}} = -i \left (
        \frac{\tilde \phi^R_{k}} {ua_k}
        -\frac{\tilde \phi^L_{j}} {ua_{j}}
    \right ).
\end{align}
We have split the operator into two terms. $\hat{\mathcal O}_{\textup{endpoints}}$ is responsible for destroying $a_{j-1}a_{j}$ quasiparticles on the $(j-1/2)$\ts{th} quasiwire and creating $a_ka_{k+1}$ quasiparticles on the $(k+1/2)$\ts{th} quasiwire. We find an additional term in \cref{eqn:theta_sum_term} which adds a phase that depends on the absolute value of $\tilde \theta$ along the path from $j$ to $k$. Since we are working on the assumption $\tilde \theta$ has well-defined expectation value, we see that the effect of this term is simply to apply a global phase of 
\begin{align}
    \gamma = \sum_{i=j}^{k-1}
    \frac{2 \braket{\tilde \theta_{i+\frac 12} (x)}}
    {u a_ia_{i+1}}
\end{align}
to the wavefunction of our system.
\begin{figure}[t]
    \centering
    \includegraphics{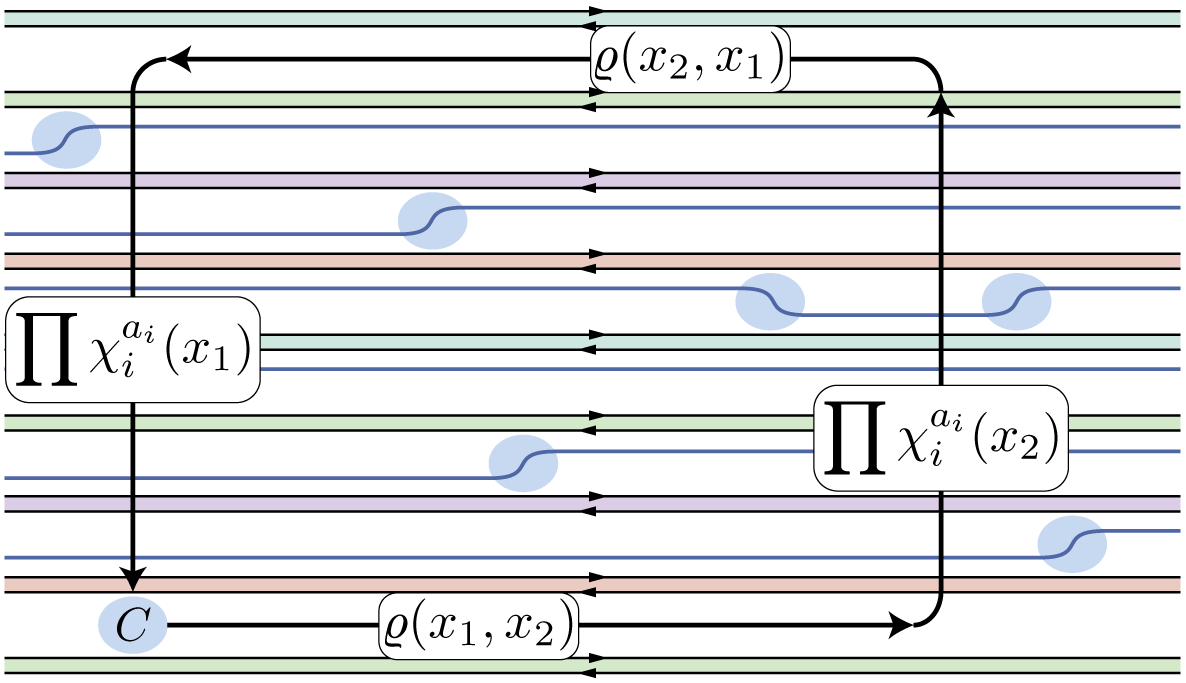}
    \caption{
        A single $C$-quasiparticle is transported around a loop, shown in black, with the corresponding operator for each section of the loop shown explicitly. Quasiparticles encircled by this loop lead to a difference in $\tilde \theta$ measured by the $C$-particle on the two sections where it moves in the $y$-direction. Thus, the overall phase accumulated on the path is a direct measure of the number of kinks contained inside the loop. 
        }
    \label{fig:loop}
\end{figure}%

Thus, let us consider moving a single $C$-type quasiparticle around a loop in the system, as shown in \cref{fig:loop}. Such an operation involves four operators. In the $x$-direction, we require two $\varrho$ operators. We work under the assumption that there are no other quasiparticles obstructing the translation of our $C$ anyon along its path---so that the phase does not precess with any non-topological phases. Thus, transport in the $x$-direction has no effect on the phase accumulated around this loop.

However, as we have just seen, transporting a kink in the $y$ direction causes the system to accumulate a phase that depends on the value of $\tilde \theta$ along the path. Thus, the full loop picks up an overall phase of 
\begin{align}\label{eqn:phase_loop}
    \gamma_{\textup{loop}}
    =
    \sum_{i=j}^{k-1}
    \frac{
        2 \braket{\tilde \theta_{i+\frac 12} (x_2)} - 
        2 \braket{\tilde \theta_{i+\frac 12} (x_1)}
        }
    {u a_ia_{i+1}}.
\end{align}
This term exactly counts the number of kinks inside the encircled region, since each single kink leads to a $\Delta \tilde \theta$ of $\pi$ between the left and right side of the loop.

We can read the mutual statistics of our anyons directly from \cref{eqn:phase_loop}. A single fractonic quasiparticle contributes a phase of $2\pi / u a_ja_{j+1}$ when encircled by a $C$-particle, whereas a $C$-type particle appears with multiplicity $a_ja_{j+1}$, and so contributes a total phase of $2\pi/u$. Thus, such C-anyons act like Laughlin quasiparticles fractionalised with the common divisor of all constituent fractions in the system. Finally, a pair of s-lineons may be braided around one another, across the only wire they are able to cross (which we choose to be the $j$\ts{th} wire for this example), incurring a phase of $2\pi/ua_j^2$---the fractionalisation of their shared wire.

\section{Fractonic Properties}\label{sec:fractonic}

As we have seen in the preceding discussion, the operators responsible for quasiparticle transport in the $y$-direction have a fracton-like constraint. For the $(j+\frac 12 )$\ts{th} quasiwire we found that the operator $\chi_j$ moves a bundle of $a_{j+1}$ quasiparticles into or out of the quasiwire, and the $\chi_{j+1}$ moves a bundle of $a_{j}$ quasiparticles. Single quasiparticles are thus immobile, since our operators cannot transport them individually. This is a feature of a fractonic phase, where excitations (in this case lineons) are constrained to move only in a lower dimensional subspace. 

If we wish to claim the system is truly fractonic, the discussion above is not sufficient \cite{gromov_fractons_2024,Nandkishore_2019}. Rather, we must demonstrate a number of properties. Firstly, we shall show in \cref{sec:constrained_fracton} that the constraint on quasiparticle transport emerges not as a consequence of our chosen transport operator ($\chi_j$), but rather as a hard constraint that \textit{any local fermionic operator must obey}. Here, a `local' operator is one that can be written in terms of bare fermionic creation operators, which themselves act locally. As a corollary, this result will imply that no perturbation can be introduced to the Hamiltonian that violates the fractonic constraints on quasiparticle configurations. 

Furthermore, a fractonic phase will generally have an exponential ground state degeneracy, since each set of lineons partially decoupled from every other and thus each contributes a multiplicative factor to the overall degeneracy. In \cref{sec:ground_state_degeneracy} we explicitly demonstrate this for our system. 

\subsection{Constrained Fractonic Transport} \label{sec:constrained_fracton}

We shall be considering a system of $N$ coupled wires with periodic boundaries in both the $x$ and $y$ directions. That is, the $0$\ts{th} wire is coupled to the $(N-1)$\ts{th} wire. Furthermore, we set the fractionalisation of each wire to $ua_j^2$, where we assume that each value of $a_j$ is coprime with $a_{j+1}$ and $a_{j-1}$ \footnote{If two adjacent wires have $a_j$ which are not coprime, we may absorb the common factor into $u$, so no new physics will arise from relaxing the assumption that adjacent $a$ are coprime.}. 

We then define the most generic fermionic operator which is written, analogous to \cref{eqn:coupling_operator}, in terms of $\psi^{L/R}$ operators. In the bosonic language, this corresponds to writing such an operator in the form $\propto e^{-i\hat {\mathcal O}}$, with
\begin{align}
    \hat {\mathcal O} = \sum_{j} 
    s_j^L\phi_j^L(x)  
    + s_j^R\phi_j^R(x). 
\end{align}
We now express the $\phi_{j}^{L/R}$ operators in terms of the quasiparticle field operators on the quasiwires above and below the $j$\ts{th} wire (the six identities used for this transformation are collected in \cref{apx:glossary_of_identities})
\begin{align}
\begin{aligned}
    \phi^{R/L}_j  =& 
    \frac {
    ua_j^2 \pm 1
    }{2}
    a_{j+1}  
    \frac{\tilde \varphi_{j+ \frac 12}}{u a_j^2 a_{j+1}^2}\\
    &+
    \frac {
    ua_j^2 \mp 1
    }{2}
    a_{j-1}  
    \frac{\tilde \varphi_{j - \frac 12}}{u a_{j-1}^2 a_{j}^2}
    + [\propto \tilde \theta].
\end{aligned}
\end{align}
Note, we have neglected to write the terms proportional to the $\tilde \theta$ fields, which do not contribute to the creation or destruction of any quasiparticles. The rescaled operators on the right-hand side of each term are precisely those that create a \textit{single} quasiparticle, since they satisfy
\begin{align}
    \left [\partial_x \tilde \theta_{j+\frac 12} (x), \frac {\tilde \varphi_{j + \frac 12}(x)}{u a_{j}^2 a_{j+1}^2} \right ] = i \pi \delta(x - x'),
\end{align}
and so place a single $\pi$-kink in the $\tilde \theta$ field on a given quasiwire. Following this, we see that the action of a $\phi_{j}^{L/R}$ is to add or remove a bundle of $(ua_j^2 \pm 1)a_{j+1}/2$ kinks from the quasiwire above, and a bundle of $(ua_j^2 \mp 1)a_{j-1}/2$ kinks from the quasiwire below, where the choice of $\pm$ is determined by whether we are applying a left or right moving bare fermion operator. 

Let us write the number of quasiparticles present on each quasiwire with an integer $n_{j+\frac 12 }$. Thus, we can express the total occupation of every wire in the system as a vector in the following way,
\begin{align}
    \textbf n = \begin{pmatrix}
        n_{0.5} &
        n_{1.5} &
        n_{2.5} &
        \cdots
        n_{N-\frac 12} 
    \end{pmatrix}^T.
\end{align}
Hence, the space of all possible number occupations of every wire is an element of $\mathbb Z^n$. The $j$\ts{th} right- or left-moving fermion creation operator acts in the following way on $\textbf n$,
\begin{align}
    n_{j+\frac 12} & \rightarrow n_{j+\frac 12} + \frac {ua_j^2 \pm 1}2 a_{j+1}, \\ 
    n_{j-\frac 12} & \rightarrow n_{j-\frac 12} + \frac {ua_j^2 \mp 1}2 a_{j-1},
\end{align}
which we may encode as in vector form as $\textbf n \rightarrow \textbf n + \textbf v^{R/L}_j$. For a system of $N$ wires, we have $2N$ fermion creation operators ($N$ right movers and $N$ left movers). We may thus express the full set of operators as an $N\times 2N$ matrix, $M$, where each column encodes the action of one of our creation operators. This matrix is written explicitly in \cref{apx:lattice_index}. The action of a \textit{completely general} local operator on $n$ can then be described with a vector $w \in \mathbb Z^{2N}$ which determines the multiplicity with which each wire has right- and left-moving fermions added (or removed),
\begin{align}\label{eqn:fracton_m_def}
    \textbf n \rightarrow
    \textbf n  + M\textbf w.
\end{align}

This operation defines a sublattice $\mathcal S \subset \mathbb Z^N$, consisting of all the configurations of quasiparticles that can be accessed by applying some collection of bare creation operators. To show that these operators obey the fractonic constraint, we must demonstrate that the successive application of local operators is not able to explore the full configuration space of $\textbf n$, which is $\mathbb Z^N$. Rather, we must show that these operators allow one to explore a (potentially infinite) subset $\mathcal S \subset \mathbb Z^N$. This problem has been extensively studied---particularly due to its relevance in post-quantum lattice-based cryptography protocols \cite{katz2018lattice}---and the proportion of sites occupied is referred to as the index of a sublattice \cite{Micciancio2002}. If the columns of $M$ provide a basis that is able to span the full extent of $\mathbb Z^N$, then the index will be $\textup{Ind} (M) = 1$, otherwise an index of $n$ will indicate that the sublattice can access $\frac 1n$ of all points in $\mathbb Z^N$. 

Determining the index of $M$ requires two steps. Firstly, our description of the lattice is overcomplete, since a lattice basis in $\mathbb Z^N$ should only require $N$ unit vectors. The first step is then to reduce our $N\times 2N$ matrix to upper triangular form using integer-Gaussian elimination over the columns. This maps $M$ onto its Hermite normal form $\tilde M$, which has non-zero elements only in an $N\times N$ upper triangular subset. The columns of $\tilde M$ still span the same sublattice, $\mathcal S$. The index of this lattice then corresponds exactly to the volume of the unit cell of $\mathcal S$, spanned by these $N$ unit vectors, since every point of $\mathbb Z^N$ within this unit cell cannot be accessed by a gauge transformation, which would only move you between unit cells. The exact calculation is laborious, and we reserve it for \cref{apx:lattice_index}. Rather, let us quote the result, which is that the index of $M$ grows exponentially in the number of wires in the system,
\begin{align}
    \textup{Ind}(M) = u\prod_{j}a_j.
\end{align}
Thus, we find that there are $u\prod_j a_j$ inequivalent configurations of fracton occupation on the wires, where {\it no local fermionic operator can move the system from one sector to another.} Starting from a configuration with zero quasiparticles, it is impossible to move it into any configuration that is not in $\mathcal S$. This constraint on fractons arises not as a consequence of any symmetry, for example charge or momentum, but rather as a consequence of locality: Any operator that breaks the fractonic constraints cannot be written in terms of bare fermionic operators, and so cannot appear in a physical system, either applied externally to move quasiparticles around, or as a perturbation to our Hamiltonian. 

\subsection{Ground State Degeneracy} \label{sec:ground_state_degeneracy}

Clearly, our system has some degree of ground state degeneracy. The only requirement to minimise the energy is that the field on each quasiwire sits at a multiple of $\pi$. Each field has an infinite number of such values that it can take. Thus, one might naively expect that the ground state has infinite degeneracy. However, some configurations of $\tilde \theta$ are gauge-equivalent. To see how, note that in bosonising our fermion fields we introduced a spurious degree of freedom. A fermionic field is unchanged under the trivial transformation $\psi \rightarrow e^{2i\pi}\psi$, so adding a multiple of $2\pi$ to the \textit{bare} bosonic fields also cannot correspond to a physical change in the system. Thus, to determine the ground state degeneracy, we must find the effect of such a gauge transformation on the quasiwire $\tilde \theta_{j+\frac 12 }$ fields. This will allow us to count the number of gauge-inequivalent configurations of the quasiwire fields, which each correspond to a single ground state. 

With this in mind, let us consider the same system as that in \cref{sec:constrained_fracton}, with $N$ coupled wires with neighbour-coprime $a_j$. On the $j$\ts{th} wire we may introduce the following two \textit{elementary gauge transformations} to the bare fields, where we add $2\pi$ either to the left moving or right moving channel,
\begin{align}
    \ \phi_{j}^R &\rightarrow \phi_{j}^R + 2\pi ,\\ 
    \phi_{j}^L &\rightarrow \phi_{j}^L + 2\pi .
\end{align}
Any arbitrary gauge transformation, adding some multiple of $2\pi$ to some chosen set of left and right movers, can be constructed by successive application of these two elementary transformations. Each of these transformations induces the following change to the quasiwire fields,
\begin{align} \label{eqn:gauge_transform_1}
    \tilde \theta_{j+\frac 12} & \rightarrow \tilde \theta_{j+\frac 12} + \frac {ua_j^2 \pm 1}2 a_{j+1}\pi, \\ \label{eqn:gauge_transform_2}
    \tilde \theta_{j-\frac 12} & \rightarrow \tilde \theta_{j-\frac 12} + \frac {ua_j^2 \mp 1}2 a_{j-1}\pi,
\end{align}
where the $\pm$ is positive for adding to the right moving field and negative for adding to the left moving field. 

For a system of $N$ wires, we have $N$ quasiwires, and the above reasoning defines a set of $2N$ elementary gauge transformations on the quasiwire fields. Thus, the naive space of all configurations is the $N$-dimensional grid, $\mathbb Z^N$, since every configuration of the fields may be represented (up to a factor of $\pi$) as an integer vector of length $N$,
\begin{align}
    \bm \Theta = \begin{pmatrix}
        \tilde \theta_{0.5} &
        \tilde \theta_{1.5} &
        \tilde \theta_{2.5} &
        \cdots
        \tilde \theta_{N-\frac 12} 
    \end{pmatrix}^T.
\end{align}
Similar to the discussion in \cref{sec:constrained_fracton}, gauge transformations then may be represented as a set of $2N$ integer vectors in this space, which we express as a $N \times 2N$ column matrix, $M$, where each column of $M$ encodes the action of one elementary gauge transformation. Remarkably, the matrix found here is identical to the matrix presented in the previous section. Any gauge transformation may be described by an arbitrary $2N$-length integer vector $\textbf w \in \mathbb Z^{2N}$ which determines the multiplicity of each elementary gauge transformation applied,
\begin{align} \label{eqn:matrix_gauge_trans}
    \bm \Theta \rightarrow \bm \Theta + M\textbf w.
\end{align}

This operation defines an identical sublattice as in the \cref{sec:constrained_fracton}, $\mathcal S \subset \mathbb Z^N$, this time representing all $\bm \Theta$ configurations which are gauge equivalent, and so can connected be with some choice of $\textbf w$ using \cref{eqn:matrix_gauge_trans}. Provided that $M$ is full rank---which we show in \cref{apx:lattice_index}---the sublattice $\mathcal S$ has dimension $N$. In this case we may define the quotient group $\mathbb Z^N / \mathcal S$, in which each element corresponds to a distinct gauge sector. This quotient group has order $\textup{Ind}(M)$, since each element represents a gauge inequivalent sublattice, which each occupy $1/\textup{Ind}(M)$ of the points in $\mathbb Z^N$. Thus, we see that the ground state degeneracy is exactly equal to the index of the lattice spanned by the matrix of gauge transformations, $M$. 
\begin{align}
    N_{G.S.} = u\prod_{j}a_j. \label{eqn:gsd}
\end{align}
Remarkably, we see that the ground state degeneracy scales exponentially with the length of the system in the $y$-direction. This is a pre-requisite for the system to have fracton excitations \cite{Nandkishore_2019}.

An important future question is to what extent the ground state degeneracy survives coupling to generic perturbations. In principle, different ground states are distinguishable if they have a configuration of $\tilde{\theta}_{j+\frac 12}$ which differ by some non-gauge amount. This means that a perturbation that can measure the different values of $\tilde{\theta}$ would lift this degeneracy, destroying the fracton order. In the case of the uniform fractional quantum Hall phase, such an operator cannot be local, and must involve a loop operator that wraps the system in the $y$-direction. Thus, in our case, the stability of the fractional phase depends on whether such a ground-state-measuring operator can be constructed locally, or scales with the system size.  

As in Ref.~\cite{santos_parafermionic_2017}, it is useful to compare our coupled-wire construction to a set of coupled clock models. Whereas they consider a single gapped boundary, and therefore a single clock model, our construction is more like a stack of one-dimensional clock models. Each individual clock model runs in the $x$ direction, with the clock model at $y=j$  having periodicity $a_j$. 
The degeneracy of such a clock model matches~\eqref{eqn:gsd} up to an overall $u$. The stacked clock model also exhibits lineon-like excitations, in the form of domain walls that can only travel in the $x$ direction. The degeneracy of the stacked clock model is not robust, because the state of each layer can be independently observed. However, as pointed out in Ref.~\cite{santos_parafermionic_2017}, the operator that measures the state of the clock model maps onto a nonlocal operator in the coupled wire construction, so that the degeneracy in the coupled wire construction is robust. Whether the clock state operator remains nonlocal in our construction is central to whether our ground state degeneracy is robust. 

Even if the ground state degeneracy is not completely robust, it can be protected by a symmetry. At the very least, there is a subsystem symmetry that relates different values of $\tilde \theta$. This symmetry is a subsystem symmetry because it acts on rigid subdimensional manifolds of the lattice. Enforcing the symmetry would lead to a symmetry-protected fracton model. A more enticing possibility is that a natural global symmetry of the physical system, such as charge conservation or translation symmetry, might protect the degeneracy. We leave the resolution of this problem to a future work.

\section{Experimental prospects}

Our results show that correlated disorder in the wire position can lead to novel fractional quantum Hall phenomenology, in sharp contrast to the known instabilities of fractional quantum Hall edges subjected to random disorder~\cite{Kane1994,Kane1995,haldane_stability_1995,Kao1999,Sheng2003,Yutushui2024}.
While we have not attempted to describe a microscopic Hamiltonian, our RG analysis (see \cref{apx:renormalisation}) shows that the couplings can be relevant. Hence, in relation to experiment,, our mechanism is relatively platform agnostic---there is a large variety of physical contexts where such couplings could dominate. 

In this section we critically describe the advantages and disadvantages of a number of state-of-the-art experimental platforms where our construction could be realised. There are two types of experiments that could realise our proposal, schematically illustrated in Fig.~\ref{fig:intro_droplets_wires}. The first type consists of those that directly realise a coupled wire construction with varying wire separation or coupling strength, see Fig.~\ref{fig:intro_droplets_wires}a. The second type arises from a system of coupled (continuum) fractional quantum Hall droplets at the desired filling fractions, see Fig.~\ref{fig:intro_droplets_wires}b.

\subsection{Solid-state Realisations}

There are several solid-state platforms that could provide a pathway towards the inhomogeneous coupled wire construction shown in Fig.~\ref{fig:intro_droplets_wires}:

\paragraph*{Coupled wires as effective models in two-dimensional materials:}
The wire construction is a realistic description for several twisted moir\'{e} heterostructures~\cite{Wu2019,Chou2019,Chen2020,Chou2021,Lee2021,Hsu2023,Biao2024,Shavit2024,Hu2024}. The wires cross in two-dimensions and are evenly spaced. Strain, ever present in real-experiments, can modify the existing couplings into inhomogeneous wire couplings. These could be solved using the approach presented in this work. However, the exact control of the inter-wire spacing is challenging.

\paragraph*{Cleaved-edge, vertically grown quantum wires:} The nanowire semiconductor industry has developed a remarkable degree of control on growing vertical nanowires. For decades it has been possible to control the size, aspect ratio, density, and crucially \textit{the position} of nanowires on multiple semiconductors~\cite{Sato2008}. Thus, the average spacing is controllable, and so is the distance between wires. The advantage is an all-electrical control but the inter-wire coupling may be weak or hard to tune precisely.
 
\paragraph*{Coupled carbon nanotube arrays:} Another possibility is to use arrays of single-wall carbon nanotubes which, when isolated, can be described as a 1D wire. Using \textit{ab-initio} methods, it has been  proposed that a planar 2D lattice of aligned carbon nanotubes can form two-dimensional networks of parallel wires, separated by an inter-wire distance of around 6\AA \cite{Polozkov2019}. However, in experiments nanotubes often overlap randomly, forming a four-fold coordinated mikado lattice~\cite{mitin2022}. While this lattice is interesting in the context of single-particle amorphous topological phases~\cite{marsal_topological_2020,marsal_obstructed_2022}, it seems challenging so far to engineer the precise inter-wire distances needed for our proposal. If future experiments can gain control of the nanotube position, for example by pinning them to a substrate, with a degree of control observed in semiconductor nanowire technology, our proposal would become feasible on this platform.

\paragraph*{Patterned quantum wires (gate-defined in 2DEG):} Standard GaAs/AlGaAs 2DEGs or two-dimensional metals like graphene~\cite{Martin2008,Laubscher2020} can be patterned by top gates or etching to define lateral quantum wires.  Arrays of parallel gate-defined wires could be engineered similar to semiconductor qubit arrays~\cite{Barthelemy2013}. 
%
Using electrodes of different sizes with electrostatic potentials of alternating sign can create effective 1D wires that can couple to one another and realise our construction~\cite{Martin2008,Laubscher2020}. 
%

\paragraph*{Coupled fractional Quantum Hall droplets:} Another plausible alternative is  to realise the coupled fractional quantum Hall droplet in Fig.~\ref{fig:intro_droplets_wires}b. Semiconductor heterostructures can realise a different fractional quantum Hall effect in neighbouring layers~\cite{Xiaomeng2019}. Experiments have also demonstrated coupling between integer and fractional quantum Hall states~\cite{Grivnin2014,Hashisaka2021,Dutta2022,Cohen2023,Hashisaka2023} where our construction is also applicable. Moreover, the recent discovery of gate tunable superconductivity, anomalous fractional and integer quantum Hall effect in MoTe$_2$ and rhombohedral graphene~\cite{xu2025signatures,Lu2025,Lu2024} opens the door to gate-engineer interfaces between different fractionalized phases proximitized by superconductors to realize our proposal.

Out of the possibilities above, the last two platforms, arrays of patterned quantum wires and Coupled fractional Quantum Hall droplets are the two most promising approaches to realise the two scenarios of our construction, Fig.~\ref{fig:intro_droplets_wires}a and b, respectively. They both offer a high degree of control of the position of the effective wires and quantum Hall droplets, and are gate tunable.

\subsection{Cold-atomic Realisations}

There are several possibilities to realise this coupled wire constructions in ultra-cold atomic gases based on existing experiments. 

\paragraph*{Coupled ultra-cold atomic tubes:} Systems based on ultra-cold atomic gases can be designed starting from arrays of 1D tubular traps. For example the inter-wire coupling can be controlled to transition between 1D to 3D phases~\cite{Revelle2016,Vogler2014,Sundar2020}. A detailed proposal is provided in Ref.~\cite{Budich2017} for realising a wire construction with one continuous and one discrete direction, along and perpendicular to the wires. The main challenge of our proposal compared to this reference is to change the confining potential $V(x,y)$ that sets the tube location. The potential is controllable externally, but may require a larger number of lasers compared to the crystalline case. To avoid this, one may tune the out-of-plane location using $V(z)$, which can be controlled independently. The last ingredient is Raman-assisted hopping~\cite{Budich2017}, which is routinely used to create an artificial flux~\cite{Cooper2019}. This strategy also has the advantage that it can realise fermionic and bosonic fractional states~\cite{Budich2017}.

\paragraph*{Few-legged Flux Ladders:} It is currently possible to realise flux ladders--- coupled wires of ultracold atoms stacked in a transverse (synthetic) direction, with an artificial magnetic flux through their plaquettes~\cite{Jaksch_2003,Celi2014,Salerno2019,crepel_microscopic_2020,Chalopin2020,Slagle2022,Zhou2023,Zhou2024,Nascimbene2025}. 
The synthetic dimension can be engineered by using different spin-states of large-spin lanthanide atoms~\cite{Chalopin2020,Nascimbene2025}, atomic momentum states~\cite{Fangzhao2018} or rotational states of molecules~\cite{Sundar2018}.
These systems are typically limited to a few wires, as the synthetic dimension is limited by the few accessible internal states. However, the coupling in the synthetic dimension is in principle highly tunable.

Out of these two possibilities, coupled ultra-cold atomic tubes separated out of the plane are the most plausible alternative to realise our proposal.

\section{Discussion and Conclusions}

In summary, we have provided an exactly solvable framework for a composite fractionalised topological order by coupling together fractional Hall phases with different fractionalisation. The framework can be used either to create new crystalline phases through stacking periodic configurations of Hall phases or non-crystalline fractional Hall phases, by stacking a non-repeating set of phases. In both cases, the model displays qualitatively different phenomenology to its uniform counterpart.

To do so, we have extended the coupled-wire construction of the fractional Hall effect by considering the properties of spatially inhomogeneous arrays of coupled wires.
Conservation of momentum requires an inhomogeneous system to also have inhomogeneous couplings, which ensure that different parts of the system have differently fractionalised excitations.
Each wire is parametrised by an integer $ua_j^2$, where $u$ defines the global fractionalisation, and each $a_j$ determines the fractionalisation around the $j$\ts{th} wire. By choosing non-uniform $a_j$, we are able to construct an inhomogeneous fractional Hall phase. 
The array of wires is coupled to form an array of quasiwires, each of which sits between two wires. At each quasiwire, the couplings can be chosen such that low-energy quasiparticles have a distinct fractionalisation, with a large freedom to choose this fractionalisation. In practice, a quasiwire between the $j$ and $j+1$ will have $ua_j^2a_{j+1}^2$-fractionalised excitations. Despite this, the system as a whole remains gapped and solvable.
We derive the exact arrangements of the wires in real space that allows for the construction to remain well-posed. This has allowed us to propose a general and solvable construction for composite and disordered, fractionalised systems.

The phenomenology of the resulting fractionalised phase is determined by the real-space arrangement of wires, and differs significantly from the uniform case.
In particular, quasiparticles in the system display a strong fractonic constraint, whereby there are completely unable to transfer between quasiwires unless grouped together into a composite excitation, motivating the description of this phase as a \textit{fractonic} fractional Hall phase. 

We find three classes of fracton excitaitions, single excitations are completely immobile, behaving as lineons. These can be grouped together to form s-lineons, which can only jump to one adjacent quasiwire and back. S-lineons can then be grouped together to form C-anyons, which are free to traverse the whole system. This restricted mobility can be understood intuitively as a consequence of different regions having different fractionalisation, so moving a single quasiparticle is forbidden as it does not correspond to a transfer of an integer amount of electron charge. The only mobile particles correspond to forming bundles of quasiparticles that have a total charge commensurate with the fractionalisation of two or more adjacent quasiwires. Futhermore, we show that this constraint is completely robust---that is, no perturbation can be introduced which allows for quasiparticle transport that breaks the fracton rules.

Further supporting the fractonic nature of the phase, we find that the ground state degeneracy depends exponentially on the number of wires in the system. An intriguing question for future work is whether this degeneracy requires the protection of a symmetry and how natural such a symmetry might be, or whether the degeneracy is robust.

Finally, we have calculated the anyons' mutual statistics, finding that the C-anyons have statistics given by the common denominator of all fractions, $u$. Single lineons (which cannot themselves braid since they are confined to a line) have mutual statistics with the composite anyons of $ua_ja_{j+1}$, for a lineon on the $(j+\frac 12)^{\textup{th}} $ quasiwire. Finally, s-lineons live adjacent to a single wire, and braid with the statistics of quasiparticles on that wire only, $ua_j^2$. Hence, both the ground state degeneracy and the excitation's properties offer richer possibilities compared to the uniform case.

We have restricted our discussion to non-uniform Laughlin fractional quantum Hall states as an example of topological order. However, similar non-uniform wire constructions can be applicable to a broad set of systems including other (potentially non-Abelian) fractionalised topological phases~\cite{Neupert2014,Klinovaja2014,oss14,sagi_non-abelian_2014,santos_fractional_2015,Klinovaja2015,Sagi2015,sagi_imprint_2015,Meng2015,meng_theory_2016,Mross2016,Iadecola2016,Fuji2016,Sagi2017,Kane2017,Park2018,Kane2018,fuji_quantum_2019,Laubscher2019,Imamura2019,Han2019,meng_coupled-wire_2020,crepel_microscopic_2020,Laubscher2020,Li_wire_2020,tam_nondiagonal_2021,Laubscher2021,Zhang2022,Laubscher2023,Pinchenkova2025}, spin liquids~\cite{meng_coupled-wire_2015,Gorohovsky2015,Patel2016,Huang2016,Lecheminant2017,Pereira2018,Ferraz2019,Slagle2022,Mondal2023,gao2025triangularlatticemodelskalmeyerlaughlin} and fractonic states~\cite{Gabor2017,Leviatan2020,sullivan_fractonic_2021,May-Mann2022,fuji_bridging_2023,You2025}.
A concrete and enticing future direction is the construction of non-abelian fractional quantum Hall states, generalizing existing results for crystalline wire constructions~\cite{teo_luttinger_2014,sagi_non-abelian_2014}. As a follow-up work, it will be interesting to bridge our findings with earlier attempts to define fractional quantum Hall states with trial wave-functions on non-crystalline lattices, such as fractals~\cite{Manna2020,Manna2022,Manna2023,Xikun2022,Jaworowski2023} and quasicrystals \cite{Duncan2020}.

\section{Acknowledgements}

We thank Jens H. Bardarson, Joan Bernabeu, Alberto Cortijo, Gwendal F\`eve, Michele Filippone, Serge Florens, Tobias Meng, C\'{e}cile Repellin, Ady Stern, and Pok Man Tam for helpful discussions. AGG and PD acknowledge financial support from the European Research Council (ERC) Consolidator grant under grant agreement No. 101042707 (TOPOMORPH). JS is supported by the program QuanTEdu-France No. ANR-22-CMAS-0001 France 2030. C.S. is supported by Vedika Khemani through the Office of Naval Research Young Investigator Program (ONR YIP) under Award Number N00014-24-1-2098.

\normalem
\bibliography{refs}

\begin{thebibliography}{211}
\bibitem [{Notei()}]{Notei}%
  \BibitemOpen
  \bibinfo {note} {Of course, for fermionic systems we require $u_j$ odd, however the same can be done starting with bosons to obtain even fractions.}\BibitemShut {Stop}%
\bibitem [{Noteii()}]{Noteii}%
  \BibitemOpen
  \bibinfo {note} {If two adjacent wires have $a_j$ which are not coprime, we may absorb the common factor into $u$, so no new physics will arise from relaxing the assumption that adjacent $a$ are coprime.}\BibitemShut {Stop}%
\end{thebibliography}

\setcounter{secnumdepth}{5}
\renewcommand{\theparagraph}{\bf \thesubsubsection.\arabic{paragraph}}

\renewcommand{\thefigure}{S\arabic{figure}}
\setcounter{figure}{0}

{\appendix

\section{Glossary of Identities}\label{apx:glossary_of_identities}
Here we present a list of several identities, used throughout the paper.

\subsection{Boson Transformations}

\begin{enumerate}[leftmargin=*, align=parleft]
    \item Sum and difference fields:
    \begin{align}
        \varphi_j &= \frac 12 \left (
            \phi^R_j + \phi^L_j
        \right ) \\
        \theta_j &= \frac 12 \left (
            \phi^R_j - \phi^L_j
        \right ) 
    \end{align}
    with inverse operation
    \begin{align}
        \phi^R_j &= \varphi_j + \theta_j \\
        \phi^L_j &= \varphi_j - \theta_j
    \end{align} 
    \item Quasiparticle Fields:
    \begin{align}
        \tilde \phi^R_j &= \varphi_j + u a_j^2 \theta_j \\
        \tilde \phi^L_j &= \varphi_j - u a_j^2 \theta_j
    \end{align}
    with inverse operation
    \begin{align} \label{eqn:apx_qp_l_inverse}
        \varphi_j &= \frac 12 \left (
            \tilde\phi^R_j + \tilde\phi^L_j
        \right ) \\ \label{eqn:apx_qp_r_inverse}
        \theta_j &= \frac 1{2ua_j^2} \left (
            \tilde\phi^R_j - \tilde\phi^L_j
        \right ) 
    \end{align}
    \item Quasiwire Sum and Difference:
    \begin{align} \label{eqn_apx_quasiwire_sum}
        \tilde\varphi_{j + \frac 12 } &= \frac 12 \left (
            a_{j+1}\tilde\phi^R_j + a_{j}\tilde\phi^L_{j+1}
        \right ) \\ \label{eqn_apx_quasiwire_diff}
        \tilde\theta_{j + \frac 12 } &= \frac 12 \left (
            a_{j+1}\tilde\phi^R_j - a_{j}\tilde\phi^L_{j+1}
        \right ) 
    \end{align}
    with inverse operation
    \begin{align}
        \tilde \phi^R_j &= 
        \frac
        {\tilde\varphi_{j + \frac 12 }+\tilde\theta_{j + \frac 12 }}
        {a_{j+1}}
         \\
        \tilde \phi^L_j &= \frac
        {\tilde\varphi_{j-\frac 12} - \tilde\theta_{j-\frac 12}}
        {a_{j-1}}
        \end{align}
\end{enumerate}

\subsection{Commutation Relations}
\begin{enumerate}[leftmargin=*, align=parleft]
    \item Bare bosonic field commutation
    \begin{align} 
        [\partial_x \phi^{L/R}_j(x), \phi^{L/R}_k(x')] = \mp 2\pi i \delta_{jk}\delta(x-x').
    \end{align}
    \item Bare sum and difference bosonic field commutation relations:
    \begin{align}
        \left [ \partial_{x} \theta_j(x) , \varphi_k(x') \right ] &=\pi i \delta_{jk} \delta(x-x'), \\
        \left [ \partial_{x} \theta_j(x) , \theta_k(x') \right ] 
        &= \left [ \partial_{x} \varphi_j(x) , \varphi_k(x') \right ] 
        = 0.
    \end{align}
    \item New bosonic fields on a wire with fractionalisation $u$,
    \begin{align} 
        [\partial_x \tilde \phi^{L/R}_j(x), \tilde \phi^{L/R}_k(x')] = \mp 2\pi i u \delta_{jk}\delta(x-x').
    \end{align}
    \item Quasiwire commutation relations where both wires have a common fraction, $u$,
    \begin{align}
         [ 
            \partial_{x} \tilde \theta_{j + \frac 12 }(x) , 
            \tilde \varphi_{k + \frac 12 }(x') 
         ] &=
        \pi i u \delta_{jk} \delta(x-x'), \\
         [
             \partial_{x} \tilde \theta_{j + \frac 12 }(x) , 
             \tilde \theta_{k + \frac 12 }(x') 
         ] 
        &=0 , \\
         [
             \partial_{x} \tilde \varphi_{j + \frac 12 }(x) , 
             \tilde \varphi_{k + \frac 12 }(x') 
         ] 
        &= 0.
    \end{align}
    \item Quasiwires with different fractions, where the $j$\ts{th} wire has fraction $ua_j^2$,
    \begin{align}
        [ \partial_{x} \tilde \theta_{j+\frac 12}(x) , 
        \tilde \varphi_{j+\frac 12}(x') ] &=
        \pi i u a_j^2a_{j+1}^2  \delta(x-x'),\label{eqn:canon_comm_tilde_1}\\
        [\partial_{x} \tilde \theta_{j+\frac 12}(x) , 
            \tilde \theta_{j+\frac 12}(x') ]&= 0 \label{eqn:canon_comm_tilde_2}\\
        [\partial_{x} \tilde \varphi_{j+\frac 12}(x) , 
        \tilde \varphi_{j+\frac 12}(x') ] 
        &= 0,\label{eqn:canon_comm_tilde_3}
    \end{align}

\end{enumerate}
\section{Charge conservation} \label{apx:substrate}

\begin{figure}[t]
    \centering
    \includegraphics{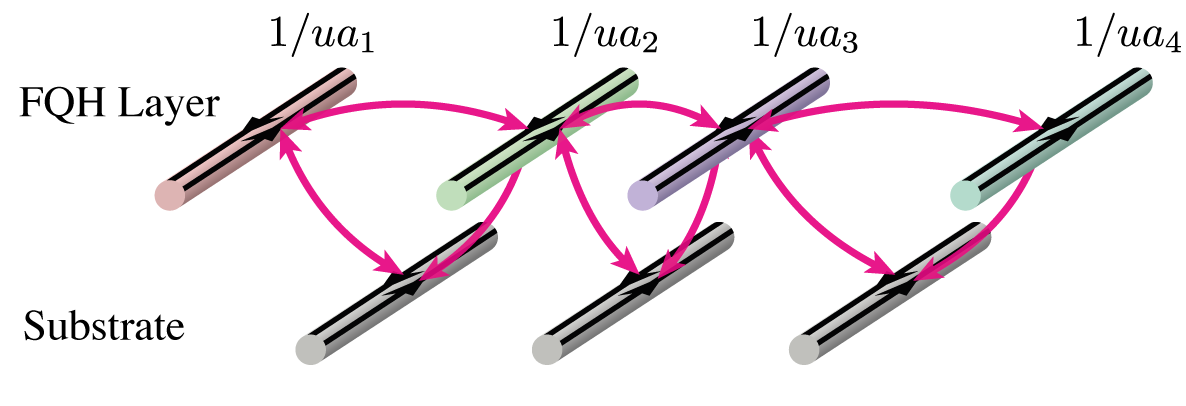}
    \caption{
    The first layer depicts a configuration of wires with each pair of wires hosting differently fractionalised quasiparticles, labelled as the fractional quantum Hall (FQH) layer. A second layer of un-fractionalised wires is added as a substrate, ensuring that all couplings respect charge conservation. 
    }
    \label{fig:apx_wires}
\end{figure}

As discussed in \cref{sec:charge_conservation}, the total charge of the system is not conserved when coupling two regions characterized by different fractional fillings~\cite{santos_parafermionic_2017, may-mann_families_2019}. 
In this Appendix, we show that coupling the system to a substrate allows us to restore charge conservation while keeping the system gapped.

Consider the coupling between two regions with fractional coefficients $ua^2$ and $ub^2$ (with $a \neq b$). The corresponding coupling operator is given by Eq.~\eqref{eq:Interface_Coupling}, which we recall here for convenience:
\begin{equation} 
\label{eqn:coupling_diff_frac} \mathcal O(x) = \cos\left ( b \varphi_1 - a\varphi_2 + ua^2b\theta_1 + uab^2\theta_2\right). 
\end{equation} 
Comparing this expression with \cref{eqn:n_m_operators}, we can extract the values of $n_1, n_2, m_1, m_2$, see \cref{eqn:n_def,eqn:m_def}. According to the charge conservation condition \cref{eqn:charge_conservation}, the net charge created or destroyed on each wire is $n_j$, and the total charge is conserved only if $n_1 + n_2 = 0$. This condition is violated here, as $(b-a)$ charges are effectively added to or removed from the system, indicating a breakdown of charge conservation.

This issue can be resolved by introducing an additional substrate wire at the interface, serving as a charge reservoir. An example configuration is shown in \cref{fig:apx_wires}. For fermions, $a$ and $b$ are odd, and hence $ a-b$ is even, allowing for the symmetric placement of $(a-b)/2$ charges on right- and left-moving modes of the extra wire. Labelling the substrate wire with the subscript $\mathrm{sub}$, we assign $s_{\mathrm{sub}}^R = (a - b)/2$ and $s_{\mathrm{sub}}^L = (a - b)/2$, resulting in:
$n_{\mathrm{sub}}=(a-b),m_{\mathrm{sub}}=0$.

The modified coupling operator, incorporating the substrate wire, becomes: 
\begin{align}
\begin{aligned} \mathcal O(x) = \cos \big( b \varphi_1 &- a\varphi_2 + (a-b)\varphi_{\mathrm{sub}}\\
         &+ ua^2b\theta_1 + uab^2\theta_2\big). 
\end{aligned} 
\label{eqn:coupling_diff_frac_EW} 
\end{align} 
Importantly, this expression only depends on the bosonic phase field $\varphi_{\mathrm{sub}}$, so the substrate wire remains non-fractionalised. The coupling always adds or removes an even number of fermions to the substrate, since $a-b$ is always even. Thus, the substrate may interpreted to be in a superconducting phase where charge is conserved modulo $a-b$~\cite{kane_fractional_2002,seroussi_topological_2014,sagi_fractional_2017}. 

To satisfy momentum conservation must we must also fulfil the condition in Eq.~\eqref{eq:momconservmain}
\begin{equation}
B\left[b y_1 - a y_2 + (a - b) y_{\mathrm{sub}}\right] + k_F u a b (a + b) = 0, 
\end{equation} 
which can be rewritten as
\begin{equation} 
B\left[b (y_1 - y_{\mathrm{sub}}) - a (y_2 - y_{\mathrm{sub}})\right] + k_F u a b (a + b) = 0. 
\end{equation} 
This condition simply reflects an origin shift, which can always be satisfied by appropriately positioning the wires.

Finally, since the field expressions associated with the fractional quantum Hall layer do not involve those of the substrate, the canonical commutation relations remain unchanged in its presence. Moreover, both the $\rho$ operator, propagating along a quasiwire and defined in ~\cref{eqn:Rho_operator}, and the $\chi$ operator, the tunnelling between neighbouring quasiwires defined in ~\cref{eqn:Chi_operator}, commute with the substrate fields and are thus unaffected by the substrate.

\section{Renormalization Group Study of Coupling Relevance} \label{apx:renormalisation}

Let us consider the relevance of the couplings proposed in \cref{sec:general_construction} under the action of renormalisation group (RG) flow. In order to demonstrate that our model is well-posed, we must show that not only are the couplings RG-relevant, but that there exists a regime in which they are the \textit{most} RG relevant operator possible. The argument takes two parts. The RG flow of couplings is determined by the kinetic term, $H_{\textup{Kin}}$, given in \cref{eqn:Main_Hkin}, where changing the form of $H_{\textup{Kin}}$ will affect which operators become relevant and which become irrelevant. Thus, we start by showing that scattering terms may be introduced that bring $H_{\textup{Kin}}$ to the correct form. Next we calculate---using this chosen kinetic term---the renormalisation equation for an arbitrary coupling. We shall find that there is a broad regime (that is, our kinetic term is not fine-tuned) under which our desired couplings have the smallest scaling dimension, making them the most relevant.

Let us start by rewriting the general form of the kinetic term for $N$ wires, written in terms of the $2N \times 2N$ matrix $M$,
\begin{align}
    H_{\textup{Kin}} = \int \frac {dx}{2\pi} 
    \begin{pmatrix}
        \partial_x \tilde {\bm \varphi} \\
        \partial_x \tilde {\bm \theta} 
    \end{pmatrix}^T
    \begin{pmatrix}
        M^{\varphi\varphi} & M^{\theta\varphi} \\
        M^{\varphi\theta} & M^{\theta\theta} 
    \end{pmatrix}
    \begin{pmatrix}
        \partial_x \tilde {\bm \varphi} \\
        \partial_x \tilde {\bm \theta} 
    \end{pmatrix}.
\end{align}
Starting with the linear bare-fermionic Hamiltonian in \cref{eqn:linearised_ham}, we will arrive at an $M$-matrix which contains non-zero off-diagonal elements---which give rise to mixed derivative terms such as  $\partial_x\tilde{\theta}_{j+\frac{1}{2}}\partial_x\tilde{\varphi}_{k+\frac{1}{2}}$. Note that for the systems we are concerned with, such mixed derivatives cannot appear. This can be seen by considering the effect of a parity transformation $\mathcal P$, which swaps $x \rightarrow -x$ and $y \rightarrow -y$. Acting on our system, such a transformation swaps $\phi^L$ and $\phi^R$, which can be followed through to our quasiwire fields,
\begin{align}
    \tilde \varphi \underset{ \mathcal P}\rightarrow \tilde \varphi,\qquad
    \tilde \theta\underset{ \mathcal P}\rightarrow -\tilde \theta.
\end{align}
Thus, for our systems, which respect parity symmetry, such mixed derivatives cannot be introduced.

However, we may incorporate forward-scattering interactions, which correspond to density-density interactions acting between channels in our system, either within a wire or between channels on different wires. Each density term has the form $\partial_x \phi^{L/R}$, which introduces a (generally rather complex) combination of $\partial_x\tilde{\theta}$ and $\partial_x\tilde{\varphi}$ to $H$. Thus, these operators give us a handle with which to tune the form of $M$. Thus, by choosing appropriate scattering terms, we can always remove the off-diagonal terms in $M$~\cite{teo_luttinger_2014,fuji_quantum_2019,tam_nondiagonal_2021}, arriving at a kinetic term of the form,
\begin{align}
    H_{\textup{Kin}} =& \sum_j\int \frac {dx}{2\pi} \left[
        M^{\theta\theta}_{j+\frac 12}
        \left (
            \partial_x\tilde{\theta}_{j + \frac 12 }
        \right )^2 + 
        M^{\varphi\varphi}_{j+\frac 12}
        \left (
            \partial_x\tilde{\varphi}_{j + \frac 12 }
        \right )^2
    \right].
\end{align}
Thus, we see that the Hamiltonian takes the form of the sine-Gordon model \cite{gianmarchi_quantum_2003}. To make this explicit, we introduce a pair of parameters 
\begin{align}
    v_{j + \frac 12 } &= \sqrt{M^{\theta\theta}_{j}M^{\varphi\varphi}_{j}}\\
    K_{j + \frac 12 } &= \sqrt{\frac{M^{\theta\theta}_{j + \frac 12 } }{M^{\varphi\varphi}_{j + \frac 12 }}},
\end{align}
arriving at the following familiar system of quasiwires,
\begin{align} \label{apx:hkin}
    H_{\textup{Kin}} =& \sum_j\int \frac {dx}{2\pi} 
    v_{j+\frac 12}
    \left[
        \frac{
         (
            \partial_x\tilde{\theta}_{j+\frac 12}
         )^2 }{K_{j+\frac 12}}
        + 
        K_{j+\frac 12}
         (
            \partial_x\tilde{\varphi}_{j+\frac 12}
         )^2
    \right].
\end{align} 

At this point, our system has decoupled into a set of separate quasiwires. Thus, let us focus on a single quasiwire, dropping the $j$ subscripts altogether, and consider adding a general coupling,
\begin{align} \label{eqn:apx_hamiltonian}
    H &= H_{\textup{Kin}} + H_{\textup{Int}},\\
        H_{\textup{Kin}} &= \int \frac {dx}{2\pi} 
    v
    \left[
        \frac{1}
        {K}
        (\partial_x\tilde{\theta})^2 
        + 
        K(\partial_x\tilde{\varphi})^2
    \right] \\
    H_{\textup{Int}} &= 
    g \int\frac{dx}{2\pi} \cos \left (
        2 \alpha \tilde \theta + 2 \beta \tilde \varphi
    \right ),
\end{align}
where $\alpha$ and $\beta$ are a pair of integer parameters that we are free to choose. Note that in order to preserve the gauge symmetry whereby $H$ is invariant under $\tilde \theta \rightarrow \tilde \theta + \pi$ (or $\tilde \varphi$), $\alpha$ and $\beta$ must be integers.  The couplings we propose in \cref{eqn:couplings_def} correspond to $(\alpha,\beta) = (1,0)$.

Our task is now to calculate the RG flow for the coupling parameter above, $g$. The calculation will follow the procedure of Appendix E of \cite{gianmarchi_quantum_2003}. First we determine the action that corresponds to the Hamiltonian given in \cref{eqn:apx_hamiltonian}. Next we shall compute correlation functions in this system. Finally, we shall use this to compute the RG flow equation for $g$.

{\it Action and Correlations---}Given the commutation relations between $\tilde \varphi$ and $\tilde \theta$, we may construct the conjugate momentum to $\tilde \theta$, $\tilde \Pi(x)$, which satisfies $[\tilde\Pi(x), \tilde \theta (x') = i \delta(x-x')]$. Comparing with the canonical commutation relations in \cref{eqn:canon_comm_tilde_1,eqn:canon_comm_tilde_2,eqn:canon_comm_tilde_3}, we find that
\begin{align}
    \tilde \Pi(x) = \frac{\partial_x \tilde \varphi(x)}{\pi m},
\end{align}
where 
\begin{align} \label{eqn:apx_m_definition}
m = u a_j^2a_{j+1}^2, 
\end{align}
encoding the fractionalisation of our chosen quasiwire.

This form of the Hamiltonian may be transformed to give the Lagrangian,
\begin{align}
    -S(\tilde \theta , \tilde \Pi) = \int d\tau \, \frac{dx}{2\pi}\, 
    \left ( 
        2\pi i  \tilde \Pi \partial_\tau \tilde \theta - H(\tilde \theta, \tilde \Pi)
    \right ).
\end{align}
Returning to a representation in terms of $\tilde \varphi$ and $\tilde \theta$, we find
\begin{align}
    -S(\tilde \theta , \tilde \varphi) =  \int d\tau  \frac{dx}{2\pi} 
    \left( 
        \frac {2i}{m}  \partial_x \tilde \varphi \partial_\tau \tilde \theta- H(\tilde \theta , \tilde \varphi)
    \right),
\end{align}
and we may write down the total partition function in terms of these fields,
\begin{align}
    \mathcal Z = \int \mathcal D \tilde \theta \mathcal D\tilde \varphi e^{
        - S(\tilde \theta , \tilde \varphi)
    }.
\end{align}
We now define $\textbf r = (v\tau, x)$ and $\textbf q = (\frac \omega v ,k)$. Let us express $ S$ in momentum space by Fourier transforming in the $x$-direction, using the following representation of the fields,
\begin{align}
    \tilde \theta (\textbf r) = \frac {1}{\beta \Omega} \sum_{\omega, k} e^{-i\textbf r \cdot \textbf q} \tilde \theta_{\textbf q},
\end{align}
with a similar expression for $\tilde \varphi (\tau,x) $. Here, $\Omega$ is the volume of the system. Thus, the kinetic part of our action takes the form,
\begin{align}
    S_{\textup{Kin}} = \frac 1{2\pi\beta \Omega} \sum_{\textbf q} \begin{pmatrix}
        \tilde \theta_{-\textbf q} \\
        \tilde \varphi_{-\textbf q} \\
    \end{pmatrix}^T 
    A_{\textbf q} \begin{pmatrix}
        \tilde \theta_{\textbf q} \\
        \tilde \varphi_{\textbf q} \\
    \end{pmatrix},
\end{align}
where $\textbf q = (\omega, vk)$, with the matrix given by,
\begin{align}
    A_{\textbf q} = \begin{pmatrix}
        \frac{vk^2} K & \frac {ik\omega }{m} \\
        \frac {ik\omega }{m}& {vk^2} K\\
    \end{pmatrix}.
\end{align}
Now we may two correlation functions that will prove useful later
\begin{align}
    \braket{\tilde \theta_{\textbf q'}\tilde \theta_{\textbf q}} &= 
    \frac 1 {\mathcal Z_0} 
    \int \mathcal D \tilde \theta \mathcal D\tilde \varphi 
    e^{
        - S_{\textup{Kin}}(\tilde \theta , \tilde \varphi)
    } \tilde \theta_{\textbf q'}\tilde \theta_{\textbf q},
    \\
    & = \beta \Omega \pi A^{-1}_{\textbf q,00}\delta_{\textbf q' -\textbf q},\\
    & = K
    \frac{
        \beta \Omega \pi v \delta_{\textbf q' -\textbf q}
    }{
        v^2 k^2 + \frac {\omega ^2}{m^2}
    }
    .\label{eqn:theta_corr}
\end{align}
Similarly, for $\tilde \varphi$, we have
\begin{align}
    \braket{\tilde \varphi_{\textbf q'}\tilde \varphi_{\textbf q}} = \frac 1 K \frac{
        \beta \Omega \pi v\delta_{\textbf q' -\textbf q}
    }{
         v^2 k^2 + \frac {\omega ^2}{m^2}
    }.\label{eqn:varphi_corr}
\end{align}

{\it Renormalisation---}We are in a position to renormalise the fields. Let us introduce a momentum cutoff $\Lambda$, along with a second cutoff at $\Lambda' < \Lambda$. We thus split our fields into fast and slow modes,
\begin{align}
    \tilde \theta (\tau, x) = 
    \tilde \theta_{-}  (\tau, x)
    + 
    \tilde \theta_{+}  (\tau, x),
\end{align}
where the separation is between Fourier modes $\tilde \theta_{-}$ consisting of states with $|\textbf q| < \Lambda'$, and $\tilde \theta_{+}$ consisting of states with $\Lambda' <|\textbf q| < \Lambda$. A similar separation is constructed for $\tilde \varphi$. Since the kinetic part of the action is diagonal in $\textbf q$, we can separate it into
$S_{\textup{Kin}} = S^{-}_{\textup{Kin}} + S^{+}_{\textup{Kin}}$. Thus, the partition function becomes,
\begin{align}
        \frac{\mathcal Z}{\mathcal Z_0} = \frac 1{\mathcal Z_0}
        \int \mathcal D \tilde \theta_- \mathcal D\tilde \varphi_-
        e^{-S_{\textup{Kin}}^-}
        \int \mathcal D \tilde \theta_+ \mathcal D\tilde \varphi_+ 
         e^{-S_{\textup{Kin}}^+ -  S^{}_{\textup{Int}}},
\end{align}
where $\mathcal Z_0$ is the partition function for $S_{\textup{Int}}=0$, which can also be separated into $\mathcal Z_0 = \mathcal Z_0^-\mathcal Z_0^+$. Our task is thus to calculate the interacting action with the high-momentum modes integrated out,
\begin{align}
    e^{-S^-_{\textup{Int}}} = \frac 1{\mathcal Z_0^+}\int \mathcal D \tilde \theta_+ \mathcal D\tilde \varphi_+ 
    e^{-S_{\textup{Kin}}^+ -  S^{}_{\textup{Int}}}.
\end{align}
We are only interested in the scaling of the coupling parameter, $g$. Thus, let us expand the interacting exponential, keeping only the terms up to first order,
\begin{align}\label{eqn:renorm_int}
        e^{-S^-_{\textup{Int}}} &= \frac 1{\mathcal Z_0^+}\int \mathcal D \tilde \theta_+ \mathcal D\tilde \varphi_+ 
        e^{-S_{\textup{Kin}}^+ }
         \left (1 -  
        S_{\textup{Int}} + ...\right ) \\
        &= 1 - \braket{S_{\textup{Int}}}_+ + ...
\end{align}
where the expectation value is an average only over fast modes. This expectation value is written explicitly as
\begin{align}
    \braket{S_{\textup{Int}}} = g \int d\tau \int\frac{dx}{2\pi} 
    \braket{
    \cos  (
        2 \alpha \tilde \theta + 2 \beta \tilde \varphi
    )}_+.
\end{align}
We may then expand this expression to the form,
\begin{align}
\begin{aligned}
    \left \langle
    \cos  (
         2 \alpha \tilde \theta + 2 \beta \tilde \varphi
    )\right \rangle_+ = &\cos (2 \alpha \tilde \theta_- + 2 \beta \tilde \varphi_-) 
    \\
    & \times e^{-2\alpha^2 \braket{\tilde \theta_+^2} - 2\beta^2 \braket{\tilde \varphi_+^2}}.
\end{aligned}
\end{align}
The justification for this expansion is tedious but straightforward. Now all that remains is to calculate $\braket{\tilde \theta_+^2(\tau, x)}$ and $\braket{\tilde \varphi_+^2(\tau, x)}$ explicitly using identities \cref{eqn:theta_corr,eqn:varphi_corr},
\begin{align}
    \braket{\tilde \theta_+^2(\tau, x)} &= 
    \frac 1 {(\beta \Omega)^2}\sum_{\Lambda ' < |\textbf q|<\Lambda }\braket{\tilde \theta_{-\textbf q}\tilde \theta_{\textbf q}}
\end{align}
Setting $\beta, L \rightarrow \infty$, we may convert this to an integral,
\begin{align}
    \braket{\tilde \theta_+^2(\tau, x)} &= 
    Kv 
    \int_{\Lambda'}^{\Lambda}  
    dk
    \int_{-\infty}^\infty 
    \frac {\pi d\omega} {v^2 k^2 + \frac {\omega^2}{m^2}} \\ 
    &= 
    Km
    \int_{\Lambda'}^{\Lambda}  
    \frac {dk} k 
    \\
    &= Km \log\left (\frac {\Lambda}{\Lambda'} \right )
\end{align}
Similarly, we find
\begin{align}
    \braket{\tilde \varphi_+^2(\tau, x)}= \frac mK \log\left (\frac {\Lambda}{\Lambda'} \right ),
\end{align}
and so the integrated coupling is given by
\begin{align}
    \braket{
    \cos  (
        2 \alpha \tilde \theta + 2 \beta \tilde \varphi
    )}_+
    = \left (\frac {\Lambda}{\Lambda'} \right )^{-\gamma} \cos (2 \alpha \tilde \theta_- + 2 \beta \tilde \varphi_-),
\end{align}
with
\begin{align}
    \gamma = 2 m\left (\alpha^2 K + \frac {\beta^2} K \right ).
\end{align}
Putting this back into the expression for the renormalised interaction, \cref{eqn:renorm_int}, we arrive at
\begin{align}
    e^{-S^-_{\textup{Int}}} = 1 - g \left (\frac {\Lambda}{\Lambda'} \right )^{-\gamma} \int\frac{d\tau\, dx}{2\pi}\cos (2 \alpha \tilde \theta_- + 2 \beta \tilde \varphi_-) + ...,
\end{align}
which we may then re-exponentiate find the renormalised interacting action,
\begin{align}
    S^-_{\textup{Int}} = g \left (\frac {\Lambda}{\Lambda'} \right )^{-\gamma} \int\frac{d\tau\, dx}{2\pi}\cos (2 \alpha \tilde \theta_- + 2 \beta \tilde \varphi_-) + \mathcal O(g^2).
\end{align}  

As the final step, we must rescale distance and time to bring the cutoff back to its original value of $\Lambda$, such that 
\begin{align}
    dx \rightarrow \frac {\Lambda}{\Lambda'} dx, \qquad d\tau \rightarrow \frac {\Lambda}{\Lambda'}d\tau.
\end{align}
Under this action, we see that the change to the coupling $g$ under our RG scheme is,
\begin{align}
    g \rightarrow  \left (\frac {\Lambda}{\Lambda'} \right )^{2-\gamma} g.
\end{align}
Thus, remembering that the $K$ value may differ between quasiwires, we return the indices from \cref{apx:hkin} and reinserting the explicit value of $m$ from \cref{eqn:apx_m_definition}. A generic coupling parametrised by $\alpha$ and $\beta$ is relevant when
\begin{align}
    \alpha^2 K_{j+\frac 12} + \frac {\beta^2} {K_{j+\frac 12}} < \frac 1{ua_j^2a_{j+1}^2}.
\end{align}
Thus, we can immediately read off two results. Firstly, our chosen coupling on each quasiwire, with $\alpha = 1, \beta = 0$ is relevant as long as 
\begin{align}
    K_{j + \frac 12}<\frac 1{ua_j^2a_{j+1}^2}.
\end{align} 
Furthermore, within this regime, any coupling with $\alpha \geq 1, \beta \geq 0$ will be strictly less relevant than the one we have chosen.

\section{The Index of a Sublattice, Fracton Constraints and Ground State Degeneracy} \label{apx:lattice_index}
In this appendix we shall explicitly calculate the index of the sublattice $\mathcal S$ encountered in \ref{sec:fractonic}. The calculation is rather tedious, since it hinges on performing partial integer Gaussian elimination on the matrix $M$ given in \cref{eqn:fracton_m_def} by hand. The result was also checked numerically and found to be consistent with our calculation. 

The index of $\mathcal S$ is defined by the following matrix,
\begin{widetext}
\begin{align}
    M = \frac 12 
    \left(\begin{array}{cccc|cccc}
    (ua_0^2 + 1) a_1 & (ua_1^2 - 1) a_0 & 0 & \cdots & (ua_0^2 - 1) a_1 & (ua_1^2 + 1) a_0 & 0 &\cdots \\
    0 & (ua_1^2 + 1) a_2 & (ua_2^2 - 1) a_1 & \cdots & 0 & (ua_1^2 - 1) a_2 &(ua_2^2 + 1) a_1 & \cdots \\
    0 & 0 & (ua_2^2 + 1) a_3 & \cdots & 0 & 0 &(ua_2^2 - 1) a_3 & \cdots \\
    \vdots & \vdots & \vdots & \ddots & \vdots & \vdots & \vdots & \ddots \\
    (ua_0^2 - 1) a_{N-1} & 0 & 0 & \cdots & (ua_0^2 + 1) a_{N-1} & 0 & 0 &\cdots 
\end{array}\right)
\end{align}
where the first $N$ columns correspond to the addition of a bare right moving fermion to each wire, and the second $N$ columns correspond to the addition of a bare left moving fermion. For clarity, we separate these two block with a vertical line. 

Calculating the index generally requires two steps. Since we have more columns, and thus more gauge transformations, than the dimenstion of the space, we must first simplify this matrix with integer Gaussian elimination. This allows one to compute the minimal lattice basis, which are given by columns of the $N\times N$ Hermite normal form, $\tilde M$, of our $N\times 2N$ matrix $M$ \cite{Micciancio2002,Cohen1993}. Once we have a reduced form, we must calculate the determinant of $\tilde M$, which will determine the ground state degeneracy.

In order to reduce the matrix we are permitted three moves: we may exchange two columns, we may add an integer multiple of a column to any other and we may multiply a column by $-1$. Thus, let us start by subtracting the right block from the left block, arriving at
\begin{align}
    \frac 12 
     \left(\begin{array}{cccc|cccc}
    2 a_1 & -2 a_0 & 0 & \cdots & (ua_0^2 - 1) a_1 & (ua_1^2 + 1) a_0 & 0 &\cdots \\
    0 & 2 a_2 & -2 a_1 & \cdots & 0 & (ua_1^2 - 1) a_2 &(ua_2^2 + 1) a_1 & \cdots \\
    0 & 0 & 2 a_3 & \cdots & 0 & 0 &(ua_2^2 - 1) a_3 & \cdots \\
    \vdots & \vdots & \vdots & \ddots & \vdots & \vdots & \vdots & \ddots\\
    -2 a_{N-1} & 0 & 0 & \cdots & (ua_0^2 + 1) a_{N-1} & 0 & 0 &\cdots 
\end{array}\right).
\end{align}
Next, we add $ (ua_0^2 + 1)/2$ copies of the 0\ts{th} column to the $N$\ts{th} column, and $(ua_1^2 + 1)/2$ copies of the 1\ts{st} column to the$(N+1)$\ts{th} column and so on. Note that since $ua_j^2$ is always odd for fermionic systems, $(a_j^2 - 1)/2$ is always an integer, so this is always possible. We arrive at the following
\begin{align}
     \left(\begin{array}{ccccc|ccccc}
     a_1 & - a_0 & 0 & \cdots & 0 & ua_0^2  a_1 & 0 & 0 &\cdots & 0
     \\
    0 &  a_2 & - a_1 & \cdots & 0 & 0 &ua_1^2  a_2 & 0 & \cdots & 0
    \\
    0 & 0 &  a_3 & \cdots & 0 & 0 & 0 &ua_2^2 a_3 & \cdots & 0
    \\
    \vdots & \vdots & \vdots & \ddots & \vdots & \vdots & \vdots & \vdots & \ddots& 0 
    \\
    0 & 0 & 0 & \cdots & -a_{N-2} & 0 & 0 & 0 & \cdots & 0
    \\
    - a_{N-1} & 0 & 0 & \cdots & a_0 & 0 & 0 & 0 &\cdots & ua_{N-1}^2 a_0
\end{array}\right),
\end{align}
where we have now included the last column of each of the two submatrices.

Let us now consider each column of the right half. Each non-zero term on the diagonal can be moved onto the row above by adding the appropriate multiple of a column from the left hand side. For example in column $N+1$ we may subtract $ua_1^2$ copies of column 1 (counting from 0) to produce a column where only the top element is non-zero, given by $ua_1^2 a_0$. We proceed with the whole right hand side matrix, arriving at the following,
\begin{align}
     \left(\begin{array}{ccccc|ccccc}
     a_1 & - a_0 & 0 & \cdots & 0 & ua_0^2  a_1 & ua_0 a_1^2  & u a_0 a_1 a_2 & u a_0 a_1 a_3 &\cdots 
     \\
    0 &  a_2 & - a_1 & \cdots & 0 & 0 & 0 & 0 & 0 & \cdots 
    \\
    0 & 0 &  a_3 & \cdots & 0 & 0 & 0 & 0 & 0 & \cdots 
    \\
    \vdots & \vdots & \vdots  & \ddots & \vdots & \vdots  & \vdots & \vdots & \vdots & \ddots
    \\
    0 & 0 & 0  & \cdots& -a_{N-2} & 0  & 0 & 0 & 0 & \cdots
    \\
    - a_{N-1} & 0 & 0 & \cdots & a_0 & 0 & 0 & 0 & 0 &\cdots 
\end{array}\right).
\end{align}
Next, we take linear combinations of the columns in the right hand side, cancelling out all columns bar one. The remaining column contains the greatest common divisor of all the entries, which is $ua_0a_1$, since adjacent $a$ are coprime, arriving at the following reduced matrix
\begin{align}
     M_{\textup{Red}} = \left(\begin{array}{ccccc|ccc}
     a_1 & - a_0 & 0 & \cdots  & 0 & u a_0  a_1 &   0 &\cdots 
     \\
    0 &  a_2 & - a_1 & \cdots & 0 & 0 & 0 & \cdots 
    \\
    0 & 0 &  a_3 & \cdots  & 0 & 0 & 0 & \cdots 
    \\
    \vdots & \vdots & \vdots & \ddots & \vdots & \vdots & \vdots & \ddots
    \\
    0 & 0 & 0 & \cdots & -a_{N-2} & 0 & 0 & 0
    \\
    - a_{N-1} & 0 & 0 & \cdots & a_0 & 0 & 0 &\cdots 
\end{array}\right).
\end{align}\end{widetext}
Thus, we have reduced the matrix to an effective $N \times N+1$ form. Despite it being trivial to reduce a {\it specific example} of this matrix to Hermite normal form numerically, the final step---removing the last column using integer Gaussian elimination---is not straightforward to do by hand in the general case. This is because it hinges on successive application of the extended Euclid's algorithm, and so depends rather non-trivially on the value of the integers $a_n$. 

Instead, we find a shortcut. The determinant of the Hermite normal form is equal to the determinant of the Smith normal form \cite{Cohen1993,tao_smith_2022}. Furthermore it has been proved that for an $n\times m$ matrix, which has rank $n$ (and consequently $m \geq n$), the determinant of the Smith normal form is given by the greatest common divisor of all full-rank minors \cite{miller_differential_2009,stanley_smith_2016}. In our case, we have a rank $N$ matrix with dimension $N\times N+1$, so must calculate its $N+1$ full-rank minors. Let us label each minor as $\delta_j$, each corresponding to the determinant of the sub-matrix formed by deleting the $j$\ts{th} column from $M_{\textup{Red}}$. Such submatrices fall into two categories:

{\it Deleting the last column:} If we remove the last column, which contains only $ua_0a_1$, the remaining matrix is singular. This can be seen by noting that the columns $\textbf c_j$ are not linearly independent. Taking the following linear combination of each column,
\begin{align}
    a_0 \textbf c_0 + a_1 \textbf c_1 + ... + a_{N-1} \textbf c_{N-1} = \bm 0,
\end{align}
we see that the matrix cannot have full rank, and thus the determinant, and corresponding minor, must vanish,
\begin{align}
    \delta_{N} = 0
\end{align}

{\it Deleting the $j$\ts{th} column of $M_{\textup{Red}}$}: Let us illustrate this first by calculating the determinant for the matrix formed by removing the first column,
\begin{align}
      \delta_{0} = \left|\begin{array}{ccccc}
     u a_0  a_1  & - a_0 & 0 & \cdots  & 0 
     \\
    0 &  a_2 & - a_1 & \cdots & 0 
    \\
    0 & 0 &  a_3 & \cdots  & 0 
    \\
    \vdots & \vdots & \vdots & \ddots & \vdots 
    \\
    0 & 0 & 0 & \cdots & a_0  
\end{array}\right|,
\end{align}
where we have reordered the columns such that the last column takes the place of the removed first column. Since this matrix is upper triangular, its determinant is expressed as the product of the diagonal elements, and the $0$\ts{th} minor is given by
\begin{align}
    \delta_0 = a_0  u \prod_j a_j.
\end{align}
For the second minor, we must find the following determinant,
\begin{align}
      \delta_{0} = \left|\begin{array}{ccccc}
     a_1  & u a_0 a_1 & 0 & \cdots  & 0 
     \\
    0 &  0 & - a_1 & \cdots & 0 
    \\
    0 & 0 &  a_3 & \cdots  & 0 
    \\
    \vdots & \vdots & \vdots & \ddots & \vdots 
    \\
    - a_{N-1} & 0 & 0 & \cdots & a_0  
\end{array}\right|.
\end{align}
The matrix is no longer upper triangular, however, note that by subtracting $u a_0$ copies of the $0$\ts{th} column to the 1\ts{st}---which does not change the determinant---we arrive at 
\begin{align}
      \delta_{0} = \left|\begin{array}{ccccc}
     a_1  & 0 & 0 & \cdots  & 0 
     \\
    0 &  0 & - a_1 & \cdots & 0 
    \\
    0 & 0 &  a_3 & \cdots  & 0 
    \\
    \vdots & \vdots & \vdots & \ddots & \vdots
    \\
    - a_{N-1} & u a_{N-1} a_0  & 0 & \cdots & a_0  
\end{array}\right|.
\end{align}
This process may be repeated until we get a matrix of the form,
\begin{align}
      \delta_{0} = \left|\begin{array}{ccccc}
     a_1  & 0 & 0 & \cdots  & 0 
     \\
    0 &  ua_1a_2 & - a_1 & \cdots & 0 
    \\
    0 & 0 &  a_3 & \cdots  & 0 
    \\
    \vdots & \vdots & \vdots & \ddots & \vdots
    \\
    - a_{N-1} & 0  & 0 & \cdots & a_0  
\end{array}\right|.
\end{align}
Finally, we shift the rows and columns by 1, which also has no effect on the determinant,
\begin{align}
      \delta_{0} = \left|\begin{array}{ccccc}
      ua_1a_2 & - a_1 & \cdots & 0 &0
    \\
     0 &  a_3 & \cdots  & 0 &0
    \\
    \vdots & \vdots  & \ddots & \vdots& \vdots 
    \\
     0  & 0 & \cdots & a_0 & - a_{N-1} 
    \\
      0 & 0 & \cdots  & 0  & a_1  
\end{array}\right|.
\end{align}
Thus, we see that the $1$\ts{st} minor is given by $\delta_1 = a_1  u \prod_j a_j$.
This may then be iterated for all remaining minors, arriving at a general expression of the form
\begin{align}
    \delta_n = a_n  u \prod_{j} a_j.
\end{align}
Since each $a_j$ is coprime to at least the two other $a$-values from its neighbouring wires, the greatest common divisor of all minors, and the index of the lattice defined by $M$ is given by
\begin{align}
    \textup{Index}(M) = u\prod_{j} a_j.
\end{align}

\end{document}